\documentclass[]{aa}  
\usepackage{graphicx}
\usepackage{txfonts}
\usepackage{lipsum}
\usepackage[dvipsnames]{xcolor}
\usepackage{stfloats}
\usepackage{multirow}
\usepackage[version=4]{mhchem}
\usepackage[utf8]{inputenc}
\usepackage{booktabs}
\usepackage{amssymb}
\usepackage{caption}

\usepackage{array}
\newcolumntype{L}[1]{>{\raggedright\let\newline\\\arraybackslash\hspace{0pt}}m{#1}}
\newcolumntype{C}[1]{>{\centering\let\newline\\\arraybackslash\hspace{0pt}}m{#1}}
\newcolumntype{R}[1]{>{\raggedleft\let\newline\\\arraybackslash\hspace{0pt}}m{#1}}
\newcommand*{\dprime}{^{\prime\prime}\mkern-1.2mu}

\definecolor{darkorange}{rgb}{1.0, 0.55, 0.0}

\usepackage{soul}

\begin{document} 
   \title{The edge-on protoplanetary disk HH~48~NE}
   \subtitle{I. Modeling the geometry and stellar parameters}
   \titlerunning{HH~48~NE: I. disk geometry}

   \author{J.A. Sturm\inst{1}\thanks{sturm@strw.leidenuniv.nl}
          \and
          M.K. McClure\inst{1}
          \and
          C.J. Law\inst{2}
          \and
          D. Harsono\inst{3}
          \and
          J.B. Bergner\inst{4}
          \and
          E. Dartois\inst{5}
          \and
          M.N. Drozdovskaya\inst{6}
          \and
          S. Ioppolo\inst{7}
          \and
          K.I. \"Oberg\inst{2}
          \and
          M.E. Palumbo\inst{8}
          \and
          Y.J. Pendleton\inst{9}
          \and
          W.R.M. Rocha\inst{1,10}
          \and
          H. Terada\inst{11,12}
          \and
          R.G. Urso\inst{8}
          }

   \institute{Leiden Observatory, Leiden University, P.O. Box 9513, NL-2300 RA Leiden, The Netherlands
    \and 
    Center for Astrophysics \textbar\ Harvard \& Smithsonian, 60 Garden St., Cambridge, MA 02138, USA
    \and
    Institute of Astronomy and Astrophysics, Academia Sinica, No. 1, Sec. 4, Roosevelt Road, Taipei 10617, Taiwan, R. O. C.
    \and
    Department of Chemistry, University of California, Berkeley, California 94720-1460, United States
    \and
    Institut des Sciences Mol\'eculaires d’Orsay, CNRS, Univ. Paris-Saclay, 91405 Orsay, France
    \and 
    Center for Space and Habitability, Universit\"at Bern, Gesellschaftsstrasse 6, 3012, Bern, Switzerland 
    \and
    Center for Interstellar Catalysis, Department of Physics and Astronomy, Aarhus University, Ny Munkegade 120, Aarhus C 8000, Denmark
    \and
    INAF – Osservatorio Astrofisico di Catania, via Santa Sofia 78, 95123 Catania, Italy
    \and 
    Department of Physics, University of Central Florida, Orlando, FL 32816, USA 
    \and
    Laboratory for Astrophysics, Leiden Observatory, Leiden University, PO Box 9513, 2300 RA Leiden, The Netherlands
    \and
    TMT International Observatory, 100 W Walnut St, Suite 300, Pasadena, CA USA
    \and
    National Astronomical Observatory of Japan, National Institutes of Natural Sciences (NINS), 2-21-1 Osawa, Mitaka, Tokyo 181-8588, Japan
    }

    \date{Received XXX; accepted YYY}
    \abstract
   {Observations of edge-on disks are an important tool for constraining general protoplanetary disk properties that cannot be determined in any other way. 
   However, most radiative transfer models cannot simultaneously reproduce the spectral energy distributions (SEDs) and resolved scattered light and submillimeter observations of these systems, due to the differences in geometry and dust properties at different wavelengths.}
   {We simultaneously constrain the geometry of the edge-on protoplanetary disk HH~48~NE and the characteristics of the host star. HH~48~NE is part of the JWST early release science program Ice Age. This work serves as a stepping stone towards a better understanding of the disk physical structure and icy chemistry in this particular source. 
   This kind of modeling lays the groundwork for studying other edge-on sources to be observed with the JWST.}
   {We fit a parameterized dust model to HH~48~NE by coupling the radiative transfer code \texttt{RADMC-3D} and an MCMC framework. The dust structure was fitted independently to a compiled SED, a scattered light image at 0.8~$\mu$m and an ALMA dust continuum observation at 890~$\mu$m.}
   {We find that 90\% of the dust mass in HH~48~NE is settled to the disk midplane, less than in average disks, and that the atmospheric layers of the disk contain exclusively large grains (0.3-10~$\mu$m).
   The exclusion of small grains in the upper atmosphere likely has important consequences for the chemistry due to the deep penetration of high-energy photons. 
   The addition of a relatively large cavity ($\sim$50 au in radius) is necessary to explain the strong mid-infrared emission, and to fit the scattered light and continuum observations simultaneously. }
   {}
   
    \maketitle
\section{Introduction}
Protoplanetary disks viewed at high inclination to the line of sight provide a unique opportunity to study the physical structures and processes that give rise to the formation of planets. 
In these systems, the cold outer disk occults both the star and the warm inner disk.
This reveals the optical and infrared emission in scattered light from small dust grains in the upper disk layers, as resolved in a limited number of Hubble Space Telescope observations \citep{Padgett1999} and ground-based telescopes.
Since the scattered light is highly sensitive to the grain properties, such observations can be used to infer grain size distributions and bulk composition \citep{Pontoppidan2005, Pontoppidan2007}. 
At submillimeter wavelengths, the vertical distributions of gas and millimeter-sized dust can be resolved with, for example, ALMA, allowing direct observations of the gas temperature structure, the CO snowline in disk atmospheres, and dynamical dust processes such as settling and radial drift \citep{Dutrey2017,Podio2020,Teague2020,Flores2021,villenave2022}. 
Combining such resolved infrared and submillimeter observations allows for the derivation of these fundamental physical quantities directly, without the integrated optical depth effects seen in face-on disks. 

Although the resolved observations may be more straightforward, modeling edge-on disks presents three unique challenges. 
First, the stellar light is blocked, which means that the stellar properties for any given disk are generally poorly constrained. 
Second, the inner disk midplane is also occulted, hiding potential radial structure that could impact the disk physical structure. 
Several studies have found suggestive evidence for hidden cavities or gaps in edge-on disks using submillimeter observations, but the impact of this on the spectral energy distribution (SED) is unclear \citep{Sauter2009, Madlener2012}.
Third, the infrared and submillimeter emission trace very different regions of the disk and are impacted by the details of the dust-gas dynamics in both the vertical and radial directions. 
This means that modeling of multi-wavelength observations including scattered light observations and resolved millimeter emission often converges to different physical models for the individual observables, or does not match with the SED \citep{wolff2017, wolff2021}. 
Furthermore, the scattered light continuum emission may involve complicated anisotropic radiative transfer from stellar radiation due to the non-trivial scattering functions, as well as mid-infrared radiation from the warm inner disk \citep{Pontoppidan2007}.

In this paper, we solve these challenges using multi-wavelength observations and modeling to find a model that reproduces all key observables. 
We focus on the edge-on protoplanetary disk HH~48~NE in the Chamaeleon~I molecular cloud, which is part of the JWST early release science program Ice Age (proposal ID: 1309, PI: McClure). 
This disk is spatially resolved by both HST \citep{Stapelfeldt2014} and ALMA \citep{Villenave2020}, allowing us to determine the disk geometry and the dynamical stellar mass.
This paper serves as a stepping stone towards understanding upcoming resolved JWST ice observations of HH~48~NE in the mid-infrared and is the first in a small series of papers. 
In this series, we combine our knowledge of the disk geometry and radiative transfer to robustly quantify future inferences possible from mid-infrared ice observations. 
In this first paper (Paper I) of the series, we introduce the source and constrain the stellar properties and disk geometry. 
In \citet[][Paper II]{PaperII}, we use the constraints from this work on the source structure to model the icy composition of HH~48~NE and determine how well we can constrain the chemistry from resolved and unresolved scattered light observations.
In later papers, we will present resolved observations and compare them with our models to ultimately measure the ice chemistry in planet-forming regions in the disk and explore what determines the chemical composition of planetary atmospheres and surfaces.

The structure of this paper is as follows: we introduce the HH~48 source and existing observations used to compare to the models in Sect.~\ref{sec:source_and_observational_data}. 
We then describe the model setup that we used and fitting procedures in Sect.~\ref{sec:model_description}. 
In Sect.~\ref{sec:results} we present the fitting results, which we discuss in detail in Sect.~\ref{sec:discussion}. 
Sect.~\ref{sec:conclusion} summarizes the results and gives final conclusions.


\begin{figure*}
\includegraphics[width=\linewidth]{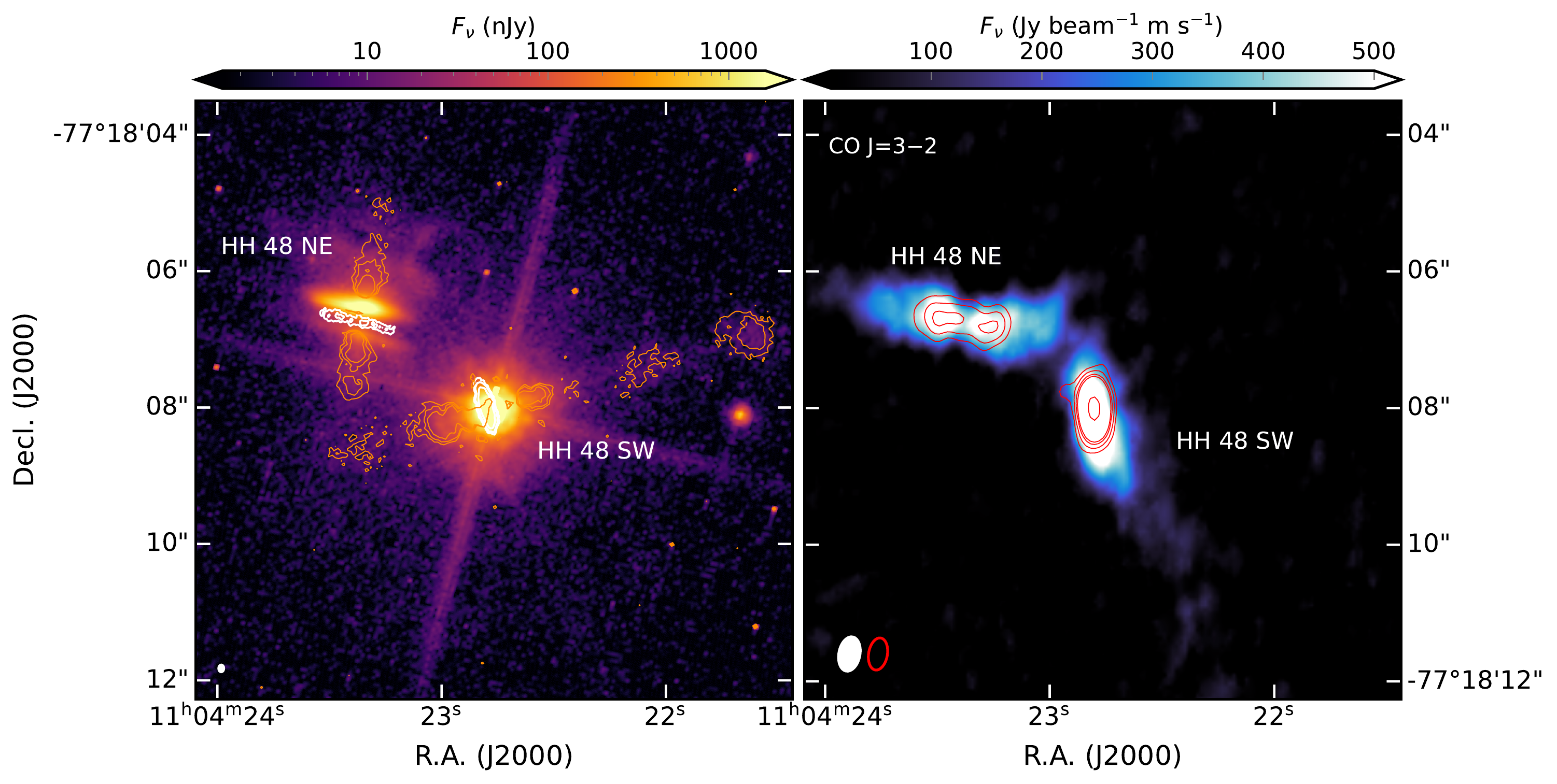}
\captionsetup{format=hang}
\caption{Overview of the HH~48 system. \\
\textbf{Left:} Scattered light observations of HH~48 with HST at 0.8~$\mu$m on a logarithmic scale to highlight the weak features. The jets observed with HST at 0.6~$\mu$m are overlayed in orange, and the ALMA Band 6 continuum (1.3~mm) is overlayed in white. The beam of the ALMA Band 6 observations (0\farcs10$\times$0\farcs07) is shown in the lower left corner.\\
\textbf{Right:} The integrated CO~$J=3-2$ emission in HH~48, with Band 7 continuum (0.89~mm) contours overlayed in red. The beams of both observations (0\farcs51$\times$0\farcs31) are shown in the lower left corner.}
\label{fig:overview}
\end{figure*}

\section{Source and observational data}
\label{sec:source_and_observational_data}
\subsection{Source description}
\label{ssec:source_description}
HH~48 (R.A.: 11$^{\rm h}$04$^{\rm m}$22.8$^{\rm s}$, Decl.: -77$^{\rm o}$18$^{\rm \prime}$08.0$\dprime$) is a binary system of protostars in the Chamaeleon I star-forming region, with a close to edge-on disk around the North-East protostar and a less inclined disk around the South-West source (see Fig.~\ref{fig:overview}). 
The separation between the two protostars is a projected distance of 2.3$\dprime$ (425 au). 
In this work, we focus on the North-East component of the binary, referred to as HH~48~NE.
HH~48 is located at a distance of 185 pc, according to recent GAIA measurements \citep{GAIA2016overview,GAIADR3} of HH~48~SW, and we assume that HH~48~NE has the same distance. 
The HH~48~NE disk appears asymmetric in scattered light observations, and has a bipolar jet that is tilted with respect to the polar axis of the disk’s midplane by 6\degr, as observed at optical wavelengths \citep[][and see Fig.~\ref{fig:overview}]{Stapelfeldt2014}. 
The disk has a sharp cut-off in millimeter continuum emission \citep[$\Delta r/r = 0.4 \pm 0.1$;][]{Villenave2020}, and shows signs of distortions in the gas disk (see Fig.~\ref{fig:overview}). 
Taken together, this likely indicates that the disks of the two young systems are interacting with each other. 

Little else is known about the system, including its central star and the protoplanetary disk. 
In the literature, only spectral types for the combined system are found, due to the small separation between the two sources.
\citet{Luhman2007} finds a K7 spectral type star, that is consistent with an effective temperature of 4000 K.
The local visual extinction ($A_{\rm v}$) in the region is estimated to be 4-6 mag, based on GAIA estimates of nearby background stars \citep{GAIADR3}.
The inclination of the disk is currently not well constrained. 
\citet{Villenave2020} finds a lower limit of $68\degr$ based on the radial and vertical extent of the disk in millimeter continuum.
No attempts at constraining the inclination from scattered light observations are found in the literature.

\subsection{Description of observations}
\label{ssec:observations}
The primary constraints on the physical properties and geometry of HH~48~NE are provided by archival HST scattered light observations, ALMA observations and the SED.
In the following sections, we introduce each of these observations.

\subsubsection{HST observations}
\label{ssec:hstobservations}
The HST observations were obtained using the Advanced Camera for Surveys (ACS) in a single orbit and are first described in \citet{Stapelfeldt2014} (GO program 12514; PI: Stapelfeldt).
For comparison with the model, we selected the exposure with the F814W filter at 0.8~$\mu$m, as this image has the best trade-off between high spatial resolution and the signal-to-noise (S/N) ratio, and is the least contaminated by the jet. 
The pipeline calibrated observations are taken from the Mikulski Archive for Space Telescopes (MAST). 
The diffuse local background in the science frame is subtracted by taking the median in an angular mask around the source at a distance of 2$\dprime$, between position angles (PAs) [-40\degr,10\degr] to avoid any contamination by the diffraction spikes and jet emission of the SW companion (see Fig. \ref{fig:overview}).
The scattered light observations are presented in Fig.~\ref{fig:overview}.

The scattered light observations reveal the two surfaces of the disk, separated with a dark lane, typical for close to edge-on protoplanetary disks. 
The intensity of the lower half of the disk is a factor of $\sim$10 weaker than of the upper half of the disk, which indicates that the source is inclined less than 90\degr. 
The disk extends radially to 1.3$\dprime$, or 240 au. 
The west side of the disk is brighter in the lower surface, which is likely a result of disk asymmetries caused by the companion.
The jet of HH~48~NE is detected in optical light (0.6~$\mu$m), but is not significantly detected at 0.8~$\mu$m. 
This is likely a result of the strong [OI] lines in the (0.6~$\mu$m) filter that trace the jet outflow. 
No direct starlight is visible at 0.8~$\mu$m, which allows us to put tight constraints on the flaring, mass and inclination of the disk.

\subsubsection{ALMA observations}
\label{ssec:almaobservations}

\subsubsection{Observational details and imaging}
We obtained ALMA archival data (2016.1.00460.S; PI: M\'enard) of the HH~48 system. 
These data consist of observations of CO~$J=3-2$ and the 0.89~mm continuum in ALMA Band 7.
See \citet{Villenave2020} for a detailed description of the observational setup and details.

The continuum observations are taken from the supplemental data products in \citet{Villenave2020}.
For the CO gas observations, we re-calibrated the data using the ALMA calibration pipeline in CASA \texttt{v5.4.0} \citep{McMullin07}.
We then self-calibrated the 0.89~mm continuum using two rounds of phase self-calibration, with solution intervals of 60s and 18s, respectively. 
The resulting calibration solutions were then applied to the full visibilities, before subtracting the continuum with a zeroth-order polynomial using the \texttt{uvcontsub} task.

We then switched to CASA \texttt{v6.3.0} for all subsequent imaging. 
We used \texttt{tclean} to produce images of CO~$J=3-2$ with a Briggs \texttt{robust} weighting of $-$2 (uniform weighting) to achieve the highest possible angular resolution. 
All images were made using the `multi-scale’ deconvolver with pixel scales of [0,5,10,25] and CLEANed down to a $4\sigma$ level, where $\sigma$ was the RMS measured from the dirty image. 
This resulted in a beam of 0\farcs51$\times$0\farcs31, PA$=-11.2^{\circ}$ for the CO~$J=3-2$ image.  
The CO~$J=3-2$ image cubes had a channel spacing of 0.21~km~s$^{-1}$ and a root mean square (RMS) value of 23~mJy~beam$^{-1}$channel$^{-1}$. 
The right panel in Fig.~\ref{fig:overview} shows the 0.89~mm continuum image and the zeroth moment maps for CO~$J=3-2$. 
The CO zeroth moment map was generated using \texttt{bettermoments} \citep{Teague18} with a $2\sigma$ clipping.
Our reprocessed continuum and that of \citet{Villenave2020} were consistent, so we just opted to use their data.

Additional continuum observations of HH~48 in Band 6 (2019.1.01792.S; PI: Mardones) were obtained from the ALMA archive.
These data were observed on 19 September 2021 at a wavelength of 1.3~mm.
The observations were carried out in two execution blocks for a total on-source integration time of 2910~s. 
In the first execution, 42 antennae were used with projected baseline lengths ranging from 15 to 3697 m. In the second execution, 34 antennae were used with baselines ranging from 47 to 8547 m.
The observations were taken in the time division mode, which means that any line emission is unresolved in the $\sim$40~km~s$^{-1}$ wide channels.
The data was calibrated using the CASA pipeline \texttt{v6.2.1} and self-calibrated using three rounds of phase-only solutions (60s, 30s, and 15s).
The continuum image was made by combining the four spectral windows with a Briggs \texttt{robust} weighting of 0.5 for a trade-off between sensitivity and angular resolution.
This resulted in a beam of 0\farcs10$\times$0\farcs07, PA$=-1.2^{\circ}$ with an RMS of 0.07~mJy~km~s$^{-1}$channel${-1}$.
The left panel in Fig.~\ref{fig:overview} shows the 1.3~mm continuum image on top of the scattered light observations. 

\begin{figure*}
    \centering
    \includegraphics[width = \linewidth]{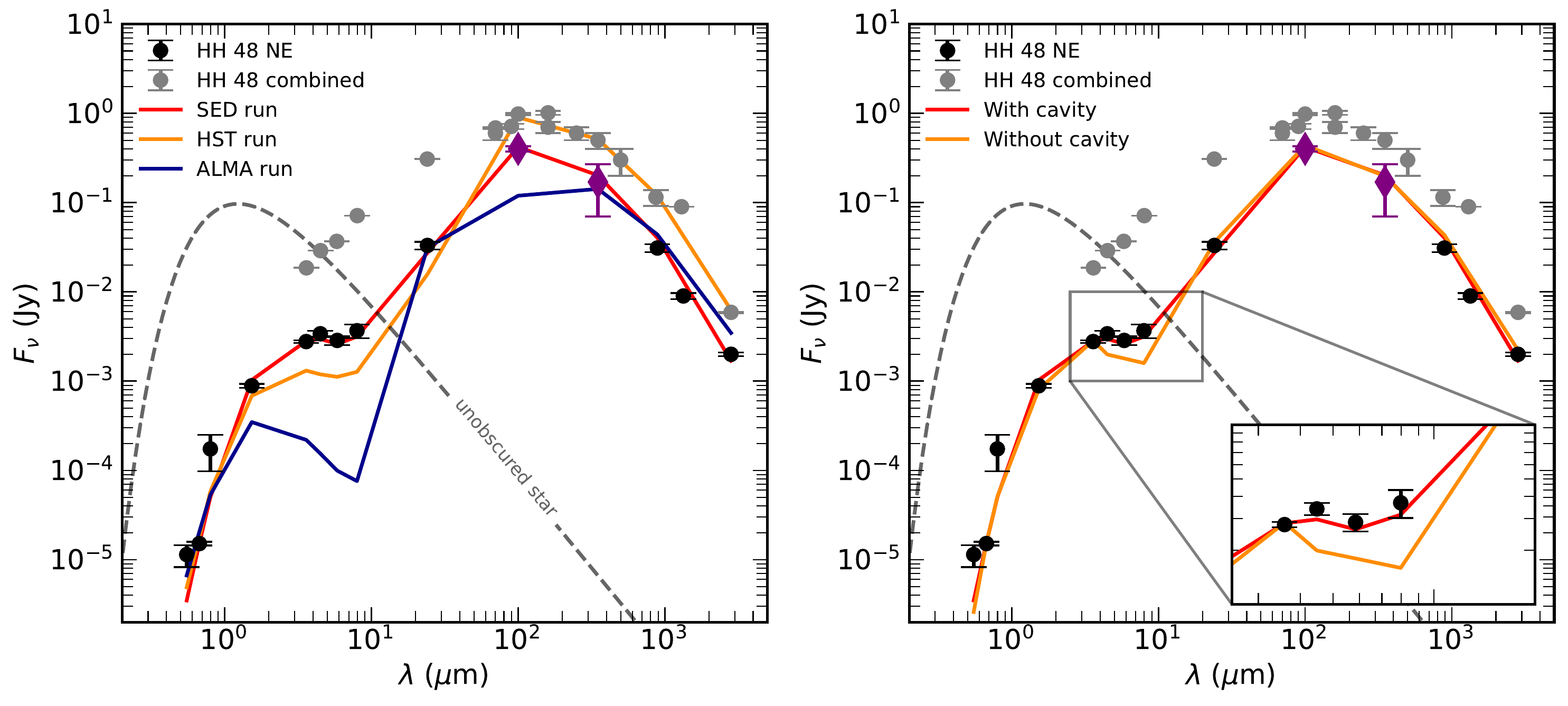}
    \captionsetup{format=hang}
    \caption{Comparison between the observed and modeled spectral energy distributions of HH~48~NE.\\
    \textbf{Left}: Observed SED of HH~48~NE (black) and observations for both of the  components of the binary system combined (grey). The flux values for HH~48~NE are listed in Table \ref{tab:Sed}. The coloured lines show the results of the three different MCMC runs described in Sect.~\ref{sec:model_description}. The dashed grey line shows the 4155 K stellar input spectrum scaled to a distance of 185 pc. The purple diamonds mark the two points that are scaled from the combined spectrum, as explained in the main text. \\ 
   \textbf{Right}: Comparison of the best fitting model in the SED runs, both with cavity and without cavity.}
    \label{fig:sed_comparison}
\end{figure*}

\subsubsection{Gas and dust morphologies}
The ALMA observations include the NE and SW components of the HH~48 system, which are both detected in CO~$J=3-2$ and continuum emission. 
The SW source is considerably brighter than the NE source in both line emission and continuum. 
HH~48~SW has a diffuse `tail' of low-intensity CO emission that extends ${\approx}$4-5$\dprime$ toward the south-west direction (see Fig.~\ref{fig:overview}). 
However, as HH~48~NE is the primary focus on this paper, further exploration of HH~48~SW is left to future work.

HH~48~NE shows a clear disk-like morphology with evidence of possible radial substructure in both continuum and CO line emission with a few suggestive emission minima and maxima visible in Fig. \ref{fig:overview}. 
In both, the 0.9 and 1.3~mm continuum observations, the disk is resolved along the radial direction with a major axis size of 1.7$\dprime$ (310 au at the distance of HH~48~NE), but is unresolved along the vertical direction.
The beam of the observations is elongated in the North-South direction due to the southern position of HH~48 on the sky.
The size of the CO gas disk extends 2.4 times further than the Band 7 continuum, with radii that enclose 95\% of the effective flux of 1.7$\dprime$ and 0.7$\dprime$, respectively. 
The CO shows a non-axisymmetric feature at the western edge of the NE source. 
At the current resolution of the ALMA observations, it is particularly difficult to discern the origin of this CO feature, which could be due to the physical disk structure (i.e., a warp), an elevated CO emitting surface \cite[e.g.,][]{Law21_MAPSIV}, or dynamical interactions between the outer edges of the NE and SW sources. 

\subsubsection{Spectral Energy Distribution}
The SED is carefully compiled from the literature using observations that resolve the NE and SW components from one another (Fig.~\ref{fig:sed_comparison}; Table \ref{tab:Sed}).
All observations with a beam larger than twice the distance between the two sources in the binary (i.e., $>4.6\dprime$) are dominated by the South West component, and should therefore be taken as upper limits. 
Observations that are included were taken by GAIA, HST, Spitzer, and ALMA. 
Unfortunately, the SED has a gap between 24 and 890~$\mu$m, where photometry of the combined system was taken at a resolution insufficient to separate the binary pair. 
No observatory is currently able to observe at this wavelength, so we are forced to scale the combined photometry to account for the NE component.
In this part of the spectrum the continuum mainly corresponds to thermal emission from the cold dust in the outer disk. 
From 890-2800 $\mu$m, the flux ratio of the two components remains approximately the same. 
Therefore, we assume the same fraction of flux (f = HH~48~NE / HH~48~combined) as the one measured at 890~$\mu$m for the 100 and 300 $\mu$m observations and scale the Herschel intensities by this factor.
The addition of these wavelength points, although with intensities including a higher error bar, are essential to discard models with SEDs significantly deviating in this wavelength range.
In Fig.~\ref{fig:sed_comparison}, we also show the observed fluxes of the two sources combined, which must not be exceeded by either of the binary components. 
Uncertainties on all data points are increased to 5\%, to account for short-term variability that often occur in young systems like these \citep{espaillat2019,zsidi2022}.

\begin{table*}[!t]

\caption{SED of HH 48 NE.}
\label{tab:Sed}
\centering
\begin{tabular}{L{.1\linewidth}L{.15\linewidth}L{.15\linewidth}L{.15\linewidth}L{.3\linewidth}}
\bottomrule
\toprule
$\lambda$ ($\mu$m) & $F_{\nu}$ (mJy)& Observatory & Angular resolution & Reference \\
\midrule
0.55 & 0.011 $\pm$ 0.003 & HST / ACS       &0.10$\dprime$ & \citet{Robberto2012}  \\
0.67 & 0.068 $\pm$ 0.006 & GAIA Band G     &0.14$\dprime$ & \citet{GAIADR3} \\
0.79 & 0.174 $\pm$ 0.009 & HST / ACS       &0.07$\dprime$ & This work\\
1.53 & 0.89  $\pm$ 0.04  & HST / WFC3      &0.13$\dprime$ & This work\\
3.59 & 2.8   $\pm$ 0.1   & Spitzer / IRAC  &0.9$\dprime$  & \citet{Dunham2016}\\
4.48 & 3.4   $\pm$ 0.3   & Spitzer / IRAC  &1.1$\dprime$  & \citet{Dunham2016}\\
5.84 & 2.9   $\pm$ 0.3   & Spitzer / IRAC  &1.4$\dprime$  & \citet{Dunham2016}\\
7.96 & 3.7   $\pm$ 0.7   & Spitzer / IRAC  &1.9$\dprime$  & \citet{Dunham2016}\\
24.1 & 33    $\pm$ 3     & Spitzer / MIPS  &5.8$\dprime$  & \citet{Dunham2016}\\
100  & 400   $\pm$ 30    & Herschel / PACS &5.9$\dprime$  & scaled from \citet{Winston2012}\\
350  & 170   $\pm$ 100   & Herschel / PACS &21$\dprime$   & scaled from \citet{Marton2017_pacs_psc}\\
890  & 31    $\pm$ 3     & ALMA            &0.4$\dprime$  & \citet{Villenave2020}\\
1339 & 7     $\pm$ 0.7   & ALMA            &0.1$\dprime$  & This work\\
2828 & 2     $\pm$ 0.1   & ALMA            &2.3$\dprime$  & \citet{Dunham2016}\\
\bottomrule
\end{tabular}

\end{table*}

\section{Modeling}
\label{sec:model_description}
\subsection{Continuum model}
Our modeling is based on the full anisotropic scattering radiative transfer capabilities of \texttt{RADMC-3D}  \citep{Dullemond2012radmc3d}. 
The specific steps of our modeling and fitting procedure are described in the following sections. 

\subsubsection{Model setup}
The model setup that we used is fully parameterized, assuming an azimuthally symmetric disk with a power law density structure and an exponential outer taper \citep{lynden-bell1974}
\begin{equation}
\label{eq:surface_density_profile}
\Sigma_{\mathrm{dust}}=\frac{\Sigma_{\mathrm{c}}}{\epsilon} \left(\frac{r}{R_{\mathrm{c}}}\right)^{-\gamma} \exp \left[-\left(\frac{r}{R_{\mathrm{c}}}\right)^{2-\gamma}\right],
\end{equation}
where $\Sigma_\mathrm{c}$ is the surface density at the characteristic radius, $R_\mathrm{c}$, $\gamma$ the power law index and $\epsilon$ the gas-to-dust ratio.
The inner radius of the disk is set to the sublimation radius, approximated by $r_{\rm subl} = 0.07\sqrt{L_{\rm s}/L_\odot}$, the outer radius of the grid is set to 300 au (1.6$\dprime$).
The stellar input spectrum is approximated by a black body with temperature ($T_{\rm s}$), scaled to the stellar luminosity ($L_{\rm s}$).

The height of the disk is described by 
\begin{equation}
\label{eq:disk_height}
h=h_{\mathrm{c}}\left(\frac{r}{R_{\mathrm{c}}}\right)^{\psi},
\end{equation}
where $h$ is the aspect ratio, $h_{\mathrm{c}}$ is the aspect ratio at the characteristic radius and $\psi$ is the flaring index.
The dust density has a vertical Gaussian distribution 
\begin{equation}
    \label{eq:small_grains_distribution}
    \rho_{\mathrm{d}}=\frac{\Sigma_{\mathrm{dust}}}{\sqrt{2 \pi} r h} \exp \left[-\frac{1}{2}\left(\frac{\pi / 2-\theta}{h}\right)^{2}\right],
\end{equation}
where $\theta$ is the opening angle from the midplane as seen from the central star. 

We adopted the canonical gas-to-dust ratio of 100 to scale between the total disk mass and the total mass in dust grains.
Dust settling is parameterized by separating the total dust mass over two dust populations, one population of small grains covering the full vertical extent of the disk, and another population with large grains that is limited in height to $X\cdot h$ with $X \in [0,1]$. 
The minimum grain size and maximum grain size of the small dust population are allowed to vary to simulate variations in grain growth. 
The minimum grain size of the large dust population is fixed to the minimum size of the small dust population, the maximum grain size of the large dust population is fixed to 1~mm. 
The two dust populations follow a power law size distribution with a fixed slope of -3.5.
The fraction of the total dust mass that resides in the large dust population is defined as $f_{\rm \ell}$.
We assumed a grain composition consistent with inter stellar matter (ISM) dust, consisting of 85\% amorphous pyroxene Mg$_{0.8}$Fe$_{0.2}$SiO$_3$, 15\% amorphous carbon and a porosity of 25\% (see e.g., \citealt{weingartner2001} and \citealt{andrews2011} for observational justification of these fixed parameters).

Dust and ice opacities were calculated using \texttt{OpTool} \citep{Dominik2021optool}, using the Distribution of Hollow Spheres \citep[DHS;][]{Min2005DHS} approach to account for grain shape effects. 
We used the full anisotropic scattering capabilities of \texttt{RADMC-3D}, as isotropic scattering assumptions can have a significant impact on the amount of observed scattered light in the near- to mid-infrared \citep{Pontoppidan2007}.

The ALMA continuum and CO gas observations suggest that the system has an inner cavity, which we discuss in more detail in Sect.~\ref{ssec:cavity}. 
To account for this in the models, we remove dust and gas inside $R_{\rm cav}$ by multiplying the surface density by a constant factor $\delta_{\rm cav}$, following the approach taken in \citet{Madlener2012}, to simulate the removal of gas and dust in the inner disk region.

In total, we have 14 free parameters that are varied to find the best fitting model for HH~48~NE. Two stellar parameters ($L_{\rm s}$, $T_{\rm s}$), six geometric parameters ($R_{\rm c}$, $h_{\rm c}$, $\psi$, $i$, $\gamma$, $M_{\rm gas}$), four parameters describing the two dust populations ($f_\ell$, $X$, $a_{\rm min}$,
$a_{\rm max}$) and two variables describing the cavity ($R_{\rm cav}$ and $\delta_{\rm cav}$).

\subsubsection{MCMC modeling}
To find a model that represents HH~48~NE the best, we applied Markov Chain Monte Carlo (MCMC) modeling on resolved observations of the disk in scattered light, the highest S/N ALMA Band 7 millimeter continuum and the SED.
MCMC is a method for sampling from probability distributions with an unknown normalization constant to converge to a global minimum in the difference between a model and observations.
Since we explored a highly degenerate, multi-parameter space, we used a parallel tempered approach \citep{Earl2005PTmcmc}. 
In parallel tempered MCMC, multiple runs sample from the probability distribution at the same time, each with a different measure (temperature) of how likely low-probability regions in the probability distribution will be crossed.
By occasionally exchanging states between these different runs, the chain of interest is less likely to remain stuck in local minima of the probability space.
We used the \texttt{python} MCMC implementation \texttt{emcee} \citep{Foreman_Mackey13}, using two different likelihood temperatures in the parallel tempered approach.

\begin{figure*}
    \centering
    \includegraphics[width = \textwidth]{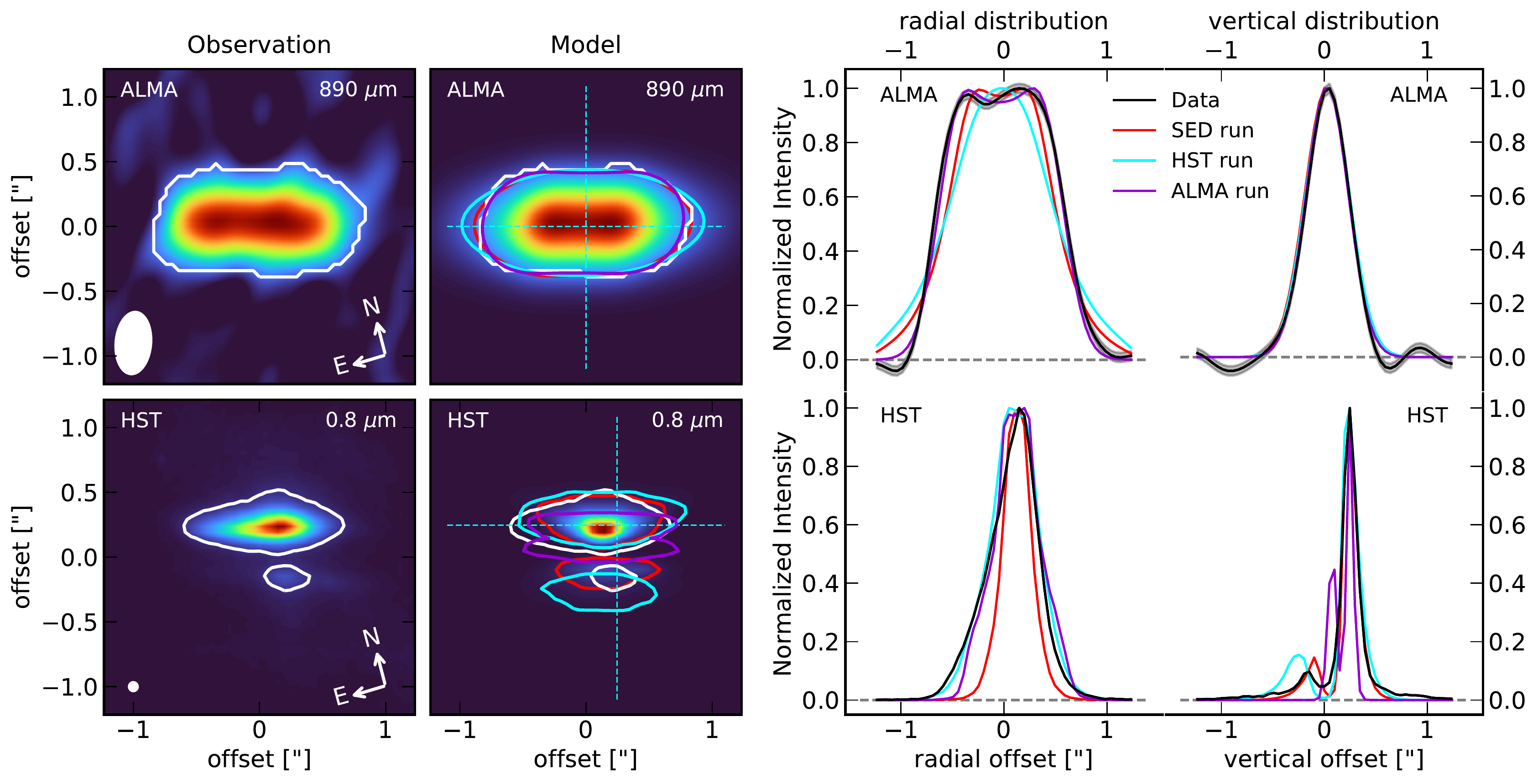}
    \captionsetup{format=hang}
    \caption{Comparison between the ALMA (890~$\mu$m, \textit{top}) and HST (0.8~$\mu$m, \textit{bottom}) observations with the ray-traced model. \\
    \textbf{Far left:} Observations of HH~48~NE normalized to the peak emission. The size of the beam/PSF is indicated in the bottom left corner. The white lines trace the 5$\sigma$ contours of the observations.\\
    \textbf{Center left:} Best fitting model at the same wavelengths as the observation, convolved with a model of the beam or PSF. The colored lines trace the same value relative to the peak flux, as the 5$\sigma$ contour of the observation in the model images for the SED run (red), the HST run (cyan) and the ALMA run (purple). The dashed, cyan line illustrates the position of the radial and vertical cuts shown in the right panel. \\
    \textbf{Center right:} Radial cut through the observations (black) and the 3 models using the SED (red), the resolved HST observation (cyan), the ALMA observation (purple) (see Sect.~\ref{sec:model_description}). The radial cuts are normalized to their respective peak flux.\\
    \textbf{Far right:} Normalized vertical cut through the data and models as for the radial distribution. The vertical cut through the HST observations is 0.25$\dprime$ off-center to get the highest S/N signal of the lower lobe.}
    \label{fig:spatial_comparison}
\end{figure*}

We used three observables to evaluate our model on:
\begin{itemize}
    \item complete SED from 0.5 - 3000~$\mu$m (see Fig.~\ref{fig:sed_comparison}),
    \item HST scattered light observation at 0.8~$\mu$m (see Fig.~\ref{fig:overview}, left panel),
    \item ALMA continuum observation at 890~$\mu$m (see Fig.~\ref{fig:overview}, right panel).
\end{itemize}
We only used the ALMA Band 7 continuum data, as this data set has the highest S/N ratio, but we compare the Band 6 continuum data extensively to the model outcome in Sect. \ref{ssec:cavity}.
Combining the different observables in one evaluation at every step in the Markov process is non-trivial, as the sources of uncertainty are different in each of them. 
Weighing the importance of the individual $\chi^2$ values is therefore not straight-forward. 
For the easiest understanding of the likelihood of each data set given by our model, we conducted three independent MCMC runs with one of the observables as input.
We will refer to each of these MCMC runs as ``SED run", ``HST run" and ``ALMA run".
An additional MCMC fit on the SED is performed without a cavity in the model, for comparison with the runs with the cavity included, and to gauge how robust the different parameters are.
We will refer to this run as ``SED run without cavity".

The models were evaluated using a $\chi^2$ value
\begin{equation}
\label{eq:chi}
\chi^{2}(\boldsymbol{m})=\sum_{i=1}^{N} \frac{\left(\mu_{i}(\boldsymbol{m})-O_{i}\right)^{2}}{\sigma_{i}^{2}}
\end{equation}
where $N$ is the number of observables, $\mu_{i}$ is the modeled observable, $O_{i}$ is the data points and $\sigma_{i}^{2}$ is the corresponding error estimate.
The log-likelihood in \texttt{emcee} is set to 
\begin{equation}
\mathcal{L}=-\frac{1}{2}\left(\chi^2+\sum_{i=1}^N \ln \left(\sigma_i^2\right)+N \ln (2 \pi)\right)
\end{equation}
\citep{wolff2017}.

For the SED runs, evaluating $\chi^2$ is straight-forward, as we can directly compare the total flux in the model with the observations and their respective uncertainty.
The modeled SEDs are corrected for foreground visual extinction. 
The $A_{\rm v}$ value is treated as an independent parameter that is determined after every run by reducing the $\chi^2_{\rm SED}$ value to a minimum assuming $A_{\rm v}$ $\geq$ 0 mag. 
Correction functions are taken from \citet{1989ApJ...345..245C} if $A_{\rm v}$ <3 mag and from \citet{McClure2009} otherwise.
For the HST and ALMA runs, we convolved the ray-traced 2D images with a Gaussian beam/point spread function (PSF) similar to that of the observations, and rebinned the images to the same resolution as the observations. 
After that we used an auto-correlation to find the optimal position of the model. 
The $\chi^2$ value is then determined over the pixels in the region where we have a 5$\sigma$ detection in the data (white contours in Fig.~\ref{fig:spatial_comparison}), to avoid over-fitting on noise features.
The images are normalized to the peak flux in both observations and models to focus only on the spatial distribution, not including the total flux at only one wavelength into the fit (as opposed to the SED run that does fit the total flux from 0.5-3000~$\mu$m).

We let 200 walkers (more than 10 times the number of free parameters) explore the parameter space for 500 steps in each of the three different runs, resulting in $2\times10^{5}$ models per run. 
A physically-motivated prior was applied to the parameter space, which means that we reject models that are not feasible given our knowledge of the general population of T Tauri stars,  to avoid long run-times and ease convergence. 
The priors are uniform distributions within the ranges given in Table~\ref{tab:prior_results}.
The disk mass, cavity depletion, grain sizes and settling fraction were sampled logarithmically to support the many orders of magnitude variations that are physically possible. 
The starting position of each of the walkers was randomly sampled using a uniform distribution in a region around a hand-picked model that fits the SED reasonably well. 
The walkers quickly spread out to sample every corner of the parameter space.
The first 100 steps were considered burn-in stage and are not taken into account for the statistics.

\begin{table*}
\caption{Results of the MCMC modeling for the four runs described in Sect.~\ref{sec:model_description}. The values in bolt are the best fitting model in that particular run, the values on the right are the median value of the distribution with the 16th and 84th percentiles of the posterior distribution as error. The evaluation ($\chi^2$ value) of the best fitting model in each run is given for each observable, with the evaluations that are not used in the optimization in italic. Note that the $A_{\rm v}$ is not sampled like the other parameters, but determined after each run by reducing $\chi^2_{\rm SED}$ to a minimum.}
\label{tab:prior_results}
\renewcommand{\arraystretch}{1.2}
\begin{tabular}{@{}L{.15\linewidth}L{.14\linewidth}L{.14\linewidth}L{.14\linewidth}L{.14\linewidth}L{.14\linewidth}@{}}
\bottomrule
\toprule
Parameter                        & Prior       & SED run                                         & HST run                                        & ALMA run                                        & SED no cavity                 \\
\midrule
$L_{\rm s}$ (L$_\odot$)          & [0.1, 3]     & \textbf{0.41}  \hfill 0.44$^{+0.11}_{-0.09}$   & \textbf{0.51} \hfill 0.48$^{+0.08}_{-0.07}$    & \textbf{0.58} \hfill 0.55$^{+0.17}_{-0.17}$     & \textbf{0.54} \hfill 0.50$^{+0.10}_{-0.07}$       \\
$T_{\rm s}$ (K)                  & [2000, 6000] & \textbf{4155}  \hfill  4500$^{+570}_{-557}$    & \textbf{4496} \hfill 4176$^{+407}_{-643}$      & \textbf{4025} \hfill 4357$^{+701}_{-764}$       & \textbf{3091} \hfill 3463$^{+593}_{-593}$         \\
\midrule
$R_{\rm c}$ (au)                 & [25, 150]    & \textbf{87}     \hfill  88$^{+29}_{-25}$       & \textbf{105}  \hfill 95$^{+13}_{-15}$          & \textbf{52} \hfill 57$^{+14}_{-13}$             & \textbf{86} \hfill 90$^{+14}_{-14}$               \\
$h_{\rm c}$                      & [0.01, 0.4]  & \textbf{0.24}  \hfill  0.23$^{+0.04}_{-0.05}$  & \textbf{0.18} \hfill 0.19$^{+0.02}_{-0.02}$    & \textbf{0.05} \hfill 0.14$^{+0.07}_{-0.05}$     & \textbf{0.21} \hfill 0.21$^{+0.05}_{-0.03}$       \\
$\psi$                           & [0.01, 0.5]  & \textbf{0.13}  \hfill  0.16$^{+0.04}_{-0.04}$  & \textbf{0.22} \hfill 0.22$^{+0.03}_{-0.03}$    & \textbf{0.25} \hfill 0.20$^{+0.07}_{-0.04}$     & \textbf{0.19} \hfill 0.20$^{+0.03}_{-0.04}$       \\
$i$ $(^{\rm o})$                 & [65, 90]     & \textbf{82.3}  \hfill  83.5$^{+3.7}_{-2.7}$    & \textbf{83.5} \hfill $83.3 \pm 1.1$            & \textbf{88.1} \hfill 88.1$^{+0.6}_{-1.2}$       & \textbf{83} \hfill 82.1$^{+2.2}_{-2.7}$           \\
$\gamma$                         & [0.3, 3]     & \textbf{0.81}  \hfill  0.77$^{+0.25}_{-0.24}$  & \textbf{0.89} \hfill 0.98$^{+0.16}_{-0.18}$    & \textbf{0.32} \hfill 0.48$^{+0.12}_{-0.10}$     & \textbf{0.93} \hfill 0.84$^{+0.20}_{-0.30}$       \\
log($M_{\rm gas}$ (M$_{\odot}$)) & [-4, -1]     & \textbf{-2.57} \hfill  -2.58$^{+0.09}_{-0.08}$ & \textbf{-1.89} \hfill -2.02$^{+0.19}_{-0.22}$  & \textbf{-2.1} \hfill -2.47$^{+0.37}_{-0.38}$    & \textbf{-2.56} \hfill -2.60$^{+0.12}_{-0.13}$     \\
\midrule
log(1-$f_{\rm \ell}$)            & [-4, 0]      & \textbf{-0.94} \hfill  -1.09$^{+0.15}_{-0.17}$ & \textbf{-0.97} \hfill -0.92$^{+0.16}_{-0.19}$  & \textbf{-1.92} \hfill -1.76$^{+0.47}_{-0.72}$   & \textbf{-1.07} \hfill -1.03$^{+0.13}_{-0.13}$     \\
$X$                              & [0.01, 0.5]  & \textbf{0.2}   \hfill  0.24$^{+0.08}_{-0.09}$  & \textbf{0.27} \hfill 0.26$^{+0.06}_{-0.05}$    & \textbf{0.47} \hfill  0.27$^{+0.08}_{-0.09}$    & \textbf{0.26} \hfill 0.26$^{+0.05}_{-0.05}$       \\
log($a_{\rm max}$ ($\mu$m))      & [-2, 1]      & \textbf{0.83}  \hfill  $0.92 \pm 0.22$         & \textbf{0.93} \hfill $0.88 \pm 0.17$           & \textbf{1.1} \hfill 1.07$^{+0.28}_{-0.22}$      & \textbf{1.0} \hfill 1.10$^{+0.29}_{-0.20}$        \\
log($a_{\rm min}$ ($\mu$m))      & [-3, 0]      & \textbf{-0.37} \hfill  -0.41$^{+0.11}_{-0.09}$ & \textbf{-0.44} \hfill -0.45$^{+0.06}_{-0.04}$  & \textbf{-0.62} \hfill -0.59$^{+0.14}_{-0.27}$   & \textbf{-0.49} \hfill -0.42$^{+0.09}_{-0.06}$     \\
\midrule
$R_{\rm cav}$ (au)               & [0, 100]     & \textbf{55}    \hfill  53$^{+14}_{-16}$        & \textbf{42.8} \hfill 52$^{+17}_{-9}$           & \textbf{80} \hfill 74$^{+5}_{-10}$              & -                             \\
log($\delta_{\rm cav}$)          & [-4, 0]      & \textbf{-1.8}  \hfill  -1.70$^{+0.47}_{-0.42}$ & \textbf{-1.76} \hfill -1.77$^{+0.19}_{-0.28}$  & \textbf{-1.5} \hfill -1.67$^{+0.24}_{-0.19}$    & -                             \\
\midrule
$\chi^2_{\rm SED}$               &              & 46                                             & \textit{3270}                                           & \textit{1109}                                            & 88                            \\
$\chi^2_{\rm HST}$               &              & \textit{25345}                                          & 8842                                           & \textit{40559}                                           & \textit{22595}                         \\
$\chi^2_{\rm ALMA}$              &              & \textit{3737}                                           & \textit{5742}                                           & 839                                             & \textit{9497}                          \\      
\midrule                                                                              
$A_{\rm v}$                      &              & 5.1                                            & 5.1                                            & 5.3                                             & 4.9                           \\
\bottomrule                        
\end{tabular}
\end{table*}

\section{Results}
\label{sec:results}
The results of the individual runs are summarized in Table~\ref{tab:prior_results}, with the uncertainties given by the 16th and 84th percentiles. 
The full posterior distributions can be found in Appendix \ref{app:posteriors}.
The MCMC runs converge to a very similar model in terms of physical parameters and geometry. 

In the forthcoming section we describe the results from the four independent MCMC runs on the SED, the HST observation in scattered light and the ALMA observation of the warm dust continuum. 
We pick one of the best fitting, representative (all parameters within 1$\sigma$) models out of each run for a comparison with the data, and discuss what each of the parameters tells us about the geometry and dust distribution of the HH~48~NE disk.
The choice of parameters for the best model in each run is indicated in Fig. \ref{fig:posteriors} and given in Table \ref{tab:prior_results} with the corresponding $\chi^2$ value on each observable. The simulated observations are compared to the data in Fig. \ref{fig:sed_comparison} and Fig. \ref{fig:spatial_comparison}.
In the ideal case, all MCMC runs end up in the same place in the parameter space, but this is unrealistic as each of the observables is sensitive to a different subset of the free parameters and the parameterized model does not account for local changes in, for example, surface density, settling, or radial drift. 
The benefit of having separate runs for the different observables, is that we can provide our own weights to the constraints of the different runs. 
In this particular case, the outcomes of the different MCMC runs are consistent with each other within the error bars for most parameters (see Table~\ref{tab:prior_results}).
The best fitting model from the SED run is taken as the overall best fit to the data and will be used as a fiducial model in the ice analysis in Paper II.
This specific model is chosen as it fits the photometric data points in the mid-infrared (see Fig. \ref{fig:sed_comparison}), which is crucial in the modeling of the ices in Paper II and the comparison with JWST data in later papers, and also reproduces the main features of the ALMA and HST observations (see Fig. \ref{fig:spatial_comparison}).

\subsection{Stellar parameters}
The best fit stellar parameters are consistent with the literature K7 spectral type for the combined system \citep{Luhman2007}. 
We find a slightly higher mean stellar effective temperature than the typical temperature of $\sim$4000 K for a K7 star \citep{Pickles1998}, but the uncertainty on the temperature is significant (10-15\%) in each of the runs.
The luminosity is well constrained at 0.4$\pm 1$~L$_\odot$, but is partly degenerate with the foreground extinction ($A_{\rm v}$) that could not be constrained from other observations due to the obscuration of the star.
The visual extinction from foreground clouds is $\sim$5 in the best fitting model, which is consistent with GAIA measurements of the extinction towards nearby stars.

\begin{figure*}
\includegraphics[trim={0 2cm 0 0},clip,width=\linewidth]{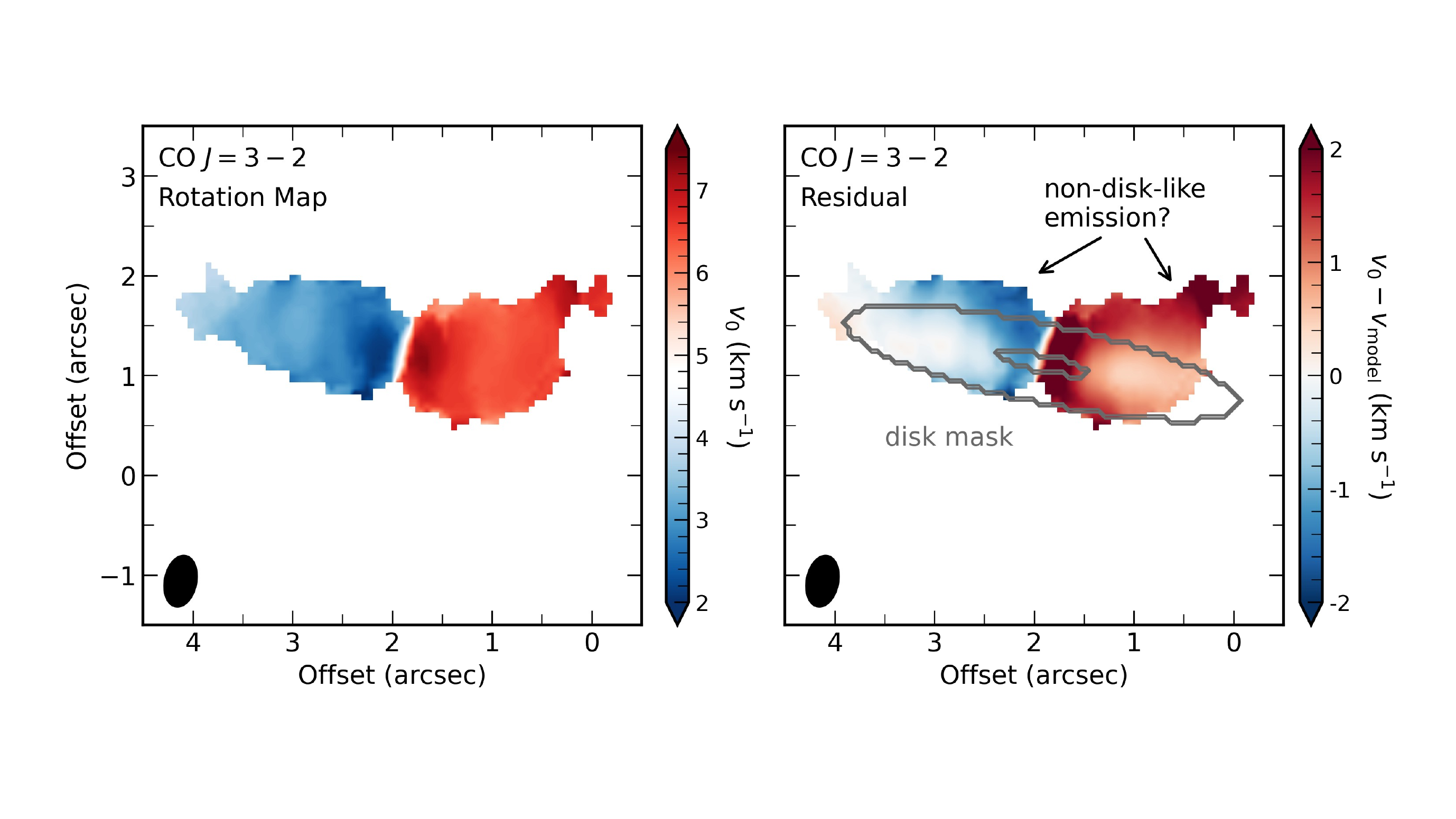}
\captionsetup{format=hang}
\caption{Dynamical mass fitting of HH~48~NE.\\
\textbf{Left:} Rotation map of the ALMA CO~$J=3-2$ observations in HH~48~NE. The spatial offset on the axes is with respect to the telescope pointing (HH~48~SW; R.A.: 166.0953\degr, Decl.: -77.30207\degr). The beam size is shown as a black ellipse in the bottom left corner (0\farcs51$\times$0\farcs31).\\
\textbf{Right:} Residuals after subtracting the model from the observations. 
The gray ellipses indicate the regions used for fitting. 
The innermost two beams were masked to avoid confusion from beam dilution and the outer radius was visually determined to avoid non-disk-like emission.
\label{fig:ALMA_mom1_fit}}
\end{figure*}

To better constrain the stellar parameters of HH~48~NE, we applied dynamical mass fitting on the CO~$J=3-2$ rotation map.
CO rotation maps can be used to derive the dynamical mass of the HH~48~NE disk, assuming that the gas is in Keplerian rotation around the center star. 
We generated a rotation map of CO~$J=3-2$ using the `quadratic’ method of \texttt{bettermoments} and excluded regions where the peak intensities are less than three times the RMS. 
Fig.~\ref{fig:ALMA_mom1_fit} shows the resulting map. 
We then fitted this rotation map with \texttt{eddy} \citep{Teague19_eddy}, which uses the \texttt{emcee} \citep{Foreman_Mackey13} python code for MCMC fitting. 
We used 64 walkers to explore the posterior distributions of the free parameters, which take 500 steps to burn in and an additional 500 steps to sample the posterior distribution function.

We consider only two free parameters in modeling the Keplerian velocity fields: stellar host mass (M$_*$) and systemic velocity (v$_{\rm{lsr}}$). 
Due to difficulties in interpreting source morphology, as noted in Sect.~\ref{ssec:almaobservations}, and the relatively large and asymmetric beam size, we fixed the coordinates of the disk center, which were visually estimated from the CO~$J=3-2$ map.
We adopted the best fit model's inclination value of 82.3\degr. 
The estimates of M$_*$ were found to be particularly sensitive to the choice of PA, which is again difficult to constrain due to uncertainties in the gas structure of HH~48~NE and the presence of significant non-Keplerian gas motions, as labeled in Fig. \ref{fig:ALMA_mom1_fit}.
We thus ran fits with a range of plausible PA values between 80-95\degr in increments of 2.5\degr. 
We note that if we adopt the PA value of 75\degr derived from the continuum in \citet{Villenave2020}, this results in a non-converging fit, again suggesting that HH~48~NE likely has a complex gas and dust structure which is only marginally resolved in the current ALMA observations.

Overall, we find best fit velocity fields consistent with a v$_{\rm{lsr}}$ of ${\approx}$4.4-4.9~km~s$^{-1}$ and a stellar host mass of ${\approx}$1--1.4~M$_{\odot}$. 
Statistical uncertainties on any individual fit are ${\approx}$20-40\%, which do not include the systematic uncertainties on each fixed parameter. 
Fig.~\ref{fig:ALMA_mom1_fit} shows a representative fit to the rotation map. 
Higher angular resolution observations of gas in the HH~48 system would greatly alleviate these difficulties and allow for a significantly more robust dynamical mass determination.

With modeled evolutionary tracks, we can check if the combination of stellar parameters that we find is theoretically feasible. 
\citet{siess2000} modeled the physical evolution of protostars using the Grenoble stellar evolution code in an extensive grid of initial conditions and list the main stellar properties as a function of time (mass, luminosity, temperature, etc.). 
Using their evolutionary tracks for solar metallicity, we find an almost perfect agreement with a luminosity of 0.4 L$_\odot$, a mass of 1 M$_\odot$, a temperature of 4070 K and an estimated age of 4-5 Myr (see Fig. \ref{fig:HRdiagram}), consistent with the age of Chameleon I.
These values are in agreement with the literature K7 spectral type, as we will discuss further in Sect.~\ref{ssec:convergence_of_fit}.

\subsection{Spatial disk parameters}\label{ssec:spatial_disk_parameters}
The disk in HH~48~NE has a characteristic radius of $\sim$90 au with a power law density index slightly less than the canonical value of 1.
The physical disk size estimate is 30\% smaller in the ALMA run than in the SED and HST runs, with a best fitting $R_\mathrm{c}$ = 57$^{+14}_{-13}$~au and $\gamma$ = 0.5 $\pm 0.1$, compared to $R_\mathrm{c}$ = 95$^{+13}_{-15}$~au and $\gamma$ = 1.0 $\pm 0.2$ using the HST observation. 
The disk appears similarly extended in the ALMA continuum image, compared to the scattered light observations (see Fig.~\ref{fig:overview}), but with a very steep outer cut-off \citep[$\Delta r/r$ = 0.4 $\pm 0.1$;][]{Villenave2020}.
In the model, we show that the disk might in fact be significantly smaller in the millimeter continuum than in the scattered light observations, taking into account radiative transfer and optical depth effects. 
This is consistent with most observations of protoplanetary disks \citep{Villenave2020}, and may be due to radial drift of the largest dust grains, or could be a result of small dust grains following the gas interaction between the two components (see Fig \ref{fig:overview}).

The disk is moderately thick with an aspect ratio, $h_\mathrm{c}$, of 0.2-0.25 and a flaring index, $\psi$, of 0.15-0.2.
These values are constrained the most by the HST run, but agree within the error bars with the SED and ALMA runs.
$h_\mathrm{c}$ and settling height $X$ are degenerate in the ALMA run, since the millimeter continuum wavelengths trace the largest grains that only exist in the large grain population that is settled with a constant $X$ with respect to the height of the small dust grains and gas.
By comparing the outcomes of the SED and HST runs, both of which trace small grains and therefore can constrain the aspect ratio better, we can conclude that the large, millimeter-sized grains are settled to 20-25\% of the disk scale height.
The mass fraction of grains that is settled to the midplane is discussed in Sect. \ref{ssec:grain_props}.

The disk is found to be inclined at 82-84\degr from the SED and HST runs.
The ALMA MCMC run suggests a higher inclination of 88\degr, but these high inclinations are clearly in conflict with the scattered light observations (see Fig.~\ref{fig:spatial_comparison}).
A similar anomaly has been seen in the extensive modeling of Oph 163131 by \citet{wolff2021}, that finds a similar offset in inclination between the scattered light observations and the ALMA continuum (which was subsequently resolved by observations at a higher angular resolution, \citealt{villenave2022}). 
Given that the ALMA image is not resolved in the vertical direction, and is only resolved with 4-5 beams along the radial direction, it is hard to break the degeneracies between the inclination and several other parameters like $h_\mathrm{c}$ and $R_\mathrm{c}$ (see Appendix \ref{app:posteriors}).
In the end, we find that an inclination of $\sim$83\degr is more probable given the appearance of the disk in scattered light, and this inclination reproduces the ALMA image relatively well (see Fig.~\ref{fig:spatial_comparison}).

\subsection{Mass}
The total mass of the disk is determined from the MCMC runs to be 2.5-10$\times 10^{-3}$ M$_\odot$, assuming a gas-to-dust ratio of 100. 
The disk mass and fraction of large grains are directly proportional with each other in the ALMA run (see Appendix \ref{app:posteriors}), as the ALMA observations only trace the large grains. 
The mass found in the ALMA run is consistent with the SED run within the systematic error when using a best fit value of $f_{\rm \ell}$ = 0.9. 
The mass estimate found based on the HST run is 4 times higher than the values found by both the ALMA and SED runs. 
High density is needed to explain the broad radial emission seen in scattered light.
The SED run fits the West wing of the upper surface in the scattered light image reasonably well (Fig.~\ref{fig:spatial_comparison}), but fails to fit the asymmetric extension on the East side of the disk.
However, the higher total disk mass is in clear conflict with the SED (see Fig.~\ref{fig:sed_comparison}), visible as an under-prediction of flux in the near- and mid-infrared and an excess of flux at millimeter wavelengths.
Therefore, the lower mass estimate is more probable. 

Using the continuum flux at 0.89~mm \citep{Villenave2020} and 2.8~mm \citep{Dunham2016}, we compare these values to the total disk mass lower limit from the observed millimeter flux using 
\begin{equation}
    M_\mathrm{gas} = 100 \frac{d^2 S_\nu }{B_\nu (T_\mathrm{d}) \kappa_\nu},
\end{equation}
where $d$ is the distance, $B_\nu (T_\mathrm{d})$ is the Planck function at the isothermal temperature $T_\mathrm{d}$, taken as 20 K, $\kappa_\nu$ is the dust opacity at the observed frequency, and the factor 100 is used to convert dust mass to total disk mass. 
Using our opacities for the large dust grain population we find an opacity of 9.1 cm$^{2}$~g$^{-1}$ and 2.3 cm$^{2}$~g$^{-1}$, corresponding to a mass estimate of 1.2$\pm$0.1$\times 10^{-3}$ M$_\odot$ and 2.4 $\pm$ 0.3$\times 10^{-3}$ M$_\odot$, for the observations at 0.89~mm and 2.8~mm, respectively.
The mass estimate at 0.89~mm is significantly lower than the estimate at 2.8~mm, which indicates that the observations at 0.89~mm are still optically thick and that we do not trace the full column of dust.
Since the column densities along the line of sight increase with inclination, this is not surprising for a disk this massive \citep[see also the discussion in][]{Villenave2020}.
The mass that we find in the MCMC runs is in agreement with these direct estimates, assuming that the bulk of the disk is optically thin at 2.8~mm.

\subsection{Grain properties}
\label{ssec:grain_props}
The minimum grain size of the small dust distribution is constrained to be 0.2-0.4~$\mu$m, excluding nm-sized dust grains in the disk atmosphere completely.
The maximum grain size of the small dust distribution is constrained to be 8-12~$\mu$m.
The small grain size distribution is best constrained by the mid-infrared part of the SED (5-100~$\mu$m) and the scattered light observations. 
The posterior distribution in the ALMA run is much broader, as the millimeter wavelength radiation is only changed via the disk temperature structure when varying the grain size distribution, rather than directly via the scattering properties. 
However, the estimates based on ALMA are in good agreement with the HST and SED runs. 

We find consistently that about 90\% of the dust mass is settled towards the midplane in large grains at a maximum height of 20-25\% of the small grains (see Sect. \ref{ssec:spatial_disk_parameters} for the discussion of the settling height).
Both the SED and the HST observations are extremely sensitive to the amount of settling, resulting in small uncertainties on the amount of small grains in the disk atmosphere. 
At inclinations <85\degr, the disk needs to be vertically extended and have a sufficient amount of small dust grains in the upper layers of the disk to block direct lines of sight or strong forward scattering from the star. 
More dust settling hence quickly results in a strong central peak in the synthesized scattered light image, that is not observed in the HST observation (see also Fig.~5 in Paper II). 

\begin{figure*}
    \includegraphics[width = \linewidth]{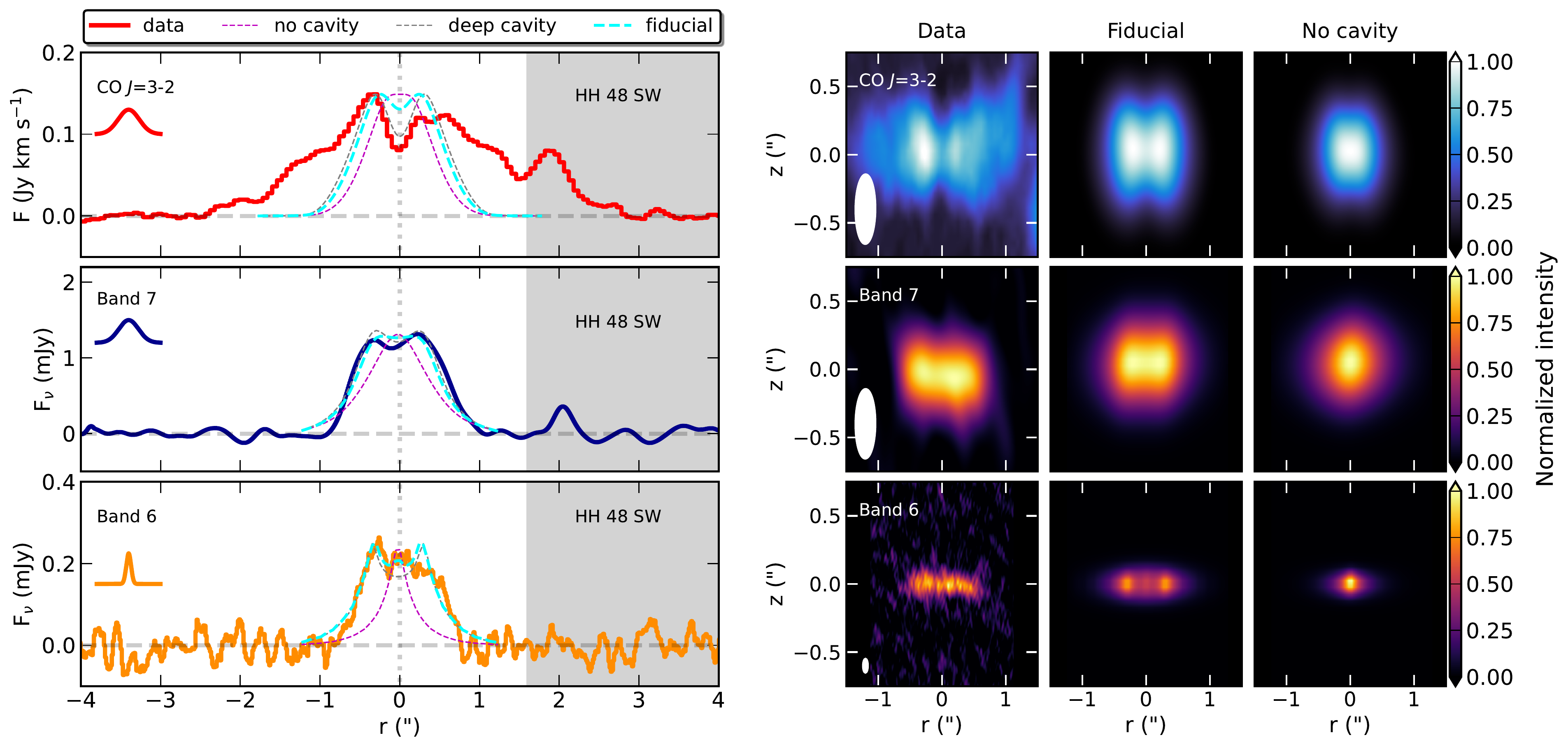}
    \captionsetup{format=hang}
    \caption{The observed radial profiles of CO and millimeter continuum from ALMA observations compared to both models with and without a cavity.\\
    \textbf{Left:} radial profile of the CO~$J=3-2$ integrated emission map (red, top), Band 7 continuum (blue, middle) and Band 6 continuum (orange, bottom). The beam size of the observations in the radial direction is illustrated with a black Gaussian profile in the upper left corner. Modeled emission is shown with a dashed line and is normalized to the peak of the observed profile. The best fitting or fiducial model is shown in cyan, the best fitting model without a cavity is shown in purple, a similar model to the best model, but with a cavity that is 100 times more depleted in gas and dust is shown in grey.
    The position of HH~48~SW is marked with a grey shaded area. \\
    \textbf{Right:} 2D maps of the CO~$J=3-2$ integrated emission (top), Band 7 continuum (middle) and Band 6 continuum (bottom). 
    Images are rotated by 12 degrees so that the disk is aligned horizontally. From left to right: the observations, the overall best fitting model and the best fitting model without a cavity. Beam sizes of the observations are shown in the lower left corner.}
    \label{fig:spatcav_evidence}
\end{figure*}

\subsection{Cavity}
\label{ssec:cavity}
We find that including a cavity in the model is necessary to fit all three observables at the same time.
To show this, we ran an additional MCMC fit on the SED without a cavity (last column in Table~\ref{tab:prior_results}) and show the SED of the resulting best fit model in Fig.~\ref{fig:sed_comparison}.

\subsubsection{SED comparison}
The modeled SEDs of the runs with and without a cavity look largely the same, but the mid-infrared flux is underestimated in the model without cavity.
The cavity decreases the amount of warm dust in the inner regions of the disk, but photons from the warm inner disk regions can escape much easier due to the significantly lower density.
These photons can then scatter again at the surface into our line of sight.
Since scattering of dust-emitted photons dominates the observed flux in the mid-infrared (2-15~$\mu$m), and this process is more efficient at longer wavelengths, where a larger part of the inner disk is optically thin, the cavity effectively boosts the flux at mid-infrared wavelengths (2-15~$\mu$m) compared to optical wavelengths (0.4-0.8~$\mu$m).
This allows combinations of parameters that do fit both the optical wavelengths and the mid-infrared wavelengths in the SED. 

The parameters of the SED without a cavity largely agree with the SED run with a cavity (see Table~\ref{tab:prior_results}), but this run prefers a significantly cooler star (3500 K) with a lower level of visual extinction to compensate for the flux deficit in the mid-infrared.
This lower temperature is in conflict with the measured dynamical mass, and is still not able to account fully for the excess flux between 2-15~$\mu$m (see Fig.~\ref{fig:sed_comparison}).

\subsubsection{Resolved continuum comparison}
Additionally, the steep radial profile with a dip in the continuum detected at 3$\sigma$ (see Fig.~\ref{fig:spatcav_evidence}) suggests that the inner region of the disk is either depleted in solids, or inclined close to 90\degr and thus optically thick \citep{Villenave2020}. 
Since the latter is excluded by the HST and SED runs, it is likely that these signatures are due to an inner cavity. 
The model of the SED run without a cavity has a strongly centrally peaked radial profile in ALMA and does not reproduce the 1$\dprime$ emission plateau and potential dip at the center (see Fig.~\ref{fig:spatcav_evidence}).
We would like to note that the effect of the cavity in the HST image is small, due to the high optical depth at visible wavelenghts.

In the ALMA Band 6 (1.3~mm) continuum, no obvious cavity is present, which sets clear limits on the size and depth of the cavity. 
Our best fitting model reproduces the size of the radial extent in the ALMA Band 6 continuum, as well as the emission plateau of $\sim$1$\dprime$. 
Models without a cavity and similar inclinations are centrally peaked and do not result in a flat profile that we observed in the ALMA Band 6 and Band 7 continuum.
Only very high inclinations would be an option \citep[see the discussion in][]{Villenave2020}, but those are excluded by the SED and scattered light observations.
As a next step, one could include an inner disk or radially dependent dust depletion factor inside the cavity in order to produce a better combined fit to the Band 6 and Band 7 data, but this is beyond the scope of this paper.

\subsubsection{Gas observation comparison}
Lastly, the CO~$J=3-2$ line emission reveals a 50\% dip at a similar position to that seen in the dust continuum (see Fig.~\ref{fig:spatcav_evidence}), and this can only be reproduced if a cavity is included or if the foreground cloud contributes significantly to the observations.
Since the foreground visual extinction is low ($A_{\rm v}$ $\sim$5 mag) and a similar dip is not observed towards HH~48~SW, the latter is unlikely.
To estimate the depth of the cavity in the gas, we added CO gas to the model at a number abundance of 10$^{-4}$ with respect to H nuclei.
To mimic freeze-out, we removed all CO gas from disk regions that are colder than 20 K, the typical desorption temperature of CO.
Photodissociation is approximated by removing all CO at gas densities lower than 10$^{7}$~cm$^{-3}$.
The CO level populations are determined in \texttt{RADMC-3D}, using a non-LTE large velocity gradient approach assuming that the gas temperature is the same as the dust temperature.
The CO~$J=3-2$ line is then ray-traced at a velocity resolution of 0.5~km~s$^{-1}$.
This simple approach is not meant to build the perfect gas model of HH~48~NE, as it has severe limitations and crude approximations, but should only be considered as a quick comparison.

We show the integrated emission of the modeled CO~$J=3-2$ line with its radial profile in Fig.~\ref{fig:spatcav_evidence}.
The radial profile of the best fitting model has a clear dip at the center, similar to the observations.
The dip is not as deep as in the observations, which could be due to the fact that we assumed the gas temperature to be the same as the dust temperature, which is no longer true in a cavity \citep{bruderer2013,Leemker2022}.
Decreasing the amount of material in the cavity by two additional orders of magnitude does reproduce the depth of the gas cavity (grey line in Fig.~\ref{fig:spatcav_evidence}), but is a less good fit to the continuum and has a significant impact on the SED.
The best fitting model without a cavity does not reproduce the observed radial profile and is strongly centrally peaked.
Hence, the gas observations show additional evidence that the disk likely has dust and gas depletion in a central region of the disk.
The modeled emission agrees with the observations in the inner 1$\dprime$, but fails to reproduce the wings that extend beyond the size of the disk in millimeter emission.
The gas in the disk extends further out than the dust, which is likely a result of radial drift \citep[see e.g.,][]{andrews2011,Trapman2019}.

\subsubsection{Size and depth}
The cavity is constrained to be 50-70 au, where dust is depleted by two orders of magnitude. 
An empty cavity results in underestimation of the mid-infrared part of the SED, so there needs to be some warm dust available in the inner cavity. 
However, it is impossible to discriminate between a depleted, yet not empty cavity or a fully depleted cavity with a small inner disk with sufficiently warm dust.
The cavity appears larger in the millimeter observations (74$^{+5}_{-10}$ au) than in the SED and HST observations ($\sim$50 au), which may be partly attributed to the poor spatial resolution of the latter observations. 
However, this may also be partly due to radial dust segregation if the small dust grains follow the gas further into the cavity than the large dust grains. 
Similar dissimilarities between the scattered light observations and millimeter continuum are seen in almost all transition disks \citep{Villenave2019,Sturm2022lkca15,vandermarel2022,Benisty2022}.

\section{Discussion}
\label{sec:discussion}
The source structure of HH~48~NE is in many ways in line with our current understanding of protoplanetary disk physics.
The height, mass and radial extent of the disk suggests that HH~48~NE has a typical T Tauri disk that would not stand out of the population of observed protoplanetary disks at lower inclination \citep[see e.g.,][]{Ansdell2016,manara2022_ppvii}.  Indeed, cavities are relatively common amongst Class II disks \citep[see e.g.,][]{vandermarel2021_deomgraphics}. 
The two main differences between this disk and the `average' disk are its relatively low mass fraction of settled large dust grains and high inclination. 
Previous infrared surveys of disks in Chamaeleon I find that between 99\% and 99.9\% of the dust mass in the disk upper layers has settled to the midplane \citep{Manoj2011}, consistent with disks in other 1-3 Myr old molecular clouds like Taurus and rho Ophiuchus \citep[][]{Furlan2011, McClure2010}. 
In contrast, only 90\% of the dust has settled in the disk around HH~48~NE. 
Given the paucity of bona fide edge-on disk detections, we speculate that the known edge-on disks could be detectable because they possess the combination of (at most) moderate dust settling and high inclination. 
We address this question in Paper II and discuss below the implications of our concordant best fitting model for the disk physical properties \citep[see also the recent paper by][]{Angelo2023}.

\subsection{Convergence of fit}
\label{ssec:convergence_of_fit}
Multi-wavelength dust fitting including the SED and resolved observations often leads to discrepancies where, for example, the scattered light observations push the model geometry in a different direction than either the continuum observations or the SED \citep[see e.g.,][]{wolff2017, wolff2021}.
Therefore, more complicated attempts, like using a staggered Markov chain \citep{Madlener2012} or covariance matrices \citep{wolff2017}, are explored in the literature, each with their promises and limitations.
We have shown that three separate MCMC runs on the SED, scattered light HST observations and mm continuum ALMA observations converge to a model that reproduces these three observables to a reasonable extent.
Additionally, we find that the temperature and luminosity in the best fitting model are in line with the measured dynamical mass of the system, given an age for the source of 4-5 Myr \citep[which is on the old end of the age estimates for the Chamaeleon I star-forming region,][]{Galli2021}. 
For that reason, we kept the procedure as simple as possible, without combining the different observations. 
As we have shown in Sect.~\ref{sec:results}, individual differences between the outputs of the runs can provide insight into the effects of dust dynamics on differently sized grains.
In principle, one could account for radial segregation of grain sizes and/or radial drift in the model, but this would introduce additional free parameters and complicate the MCMC fits while the data may not have high enough S/N and/or spatial resolution to definitively draw conclusions on the differences between large and small grains.

\begin{figure}[ht]
    \includegraphics[width = \linewidth]{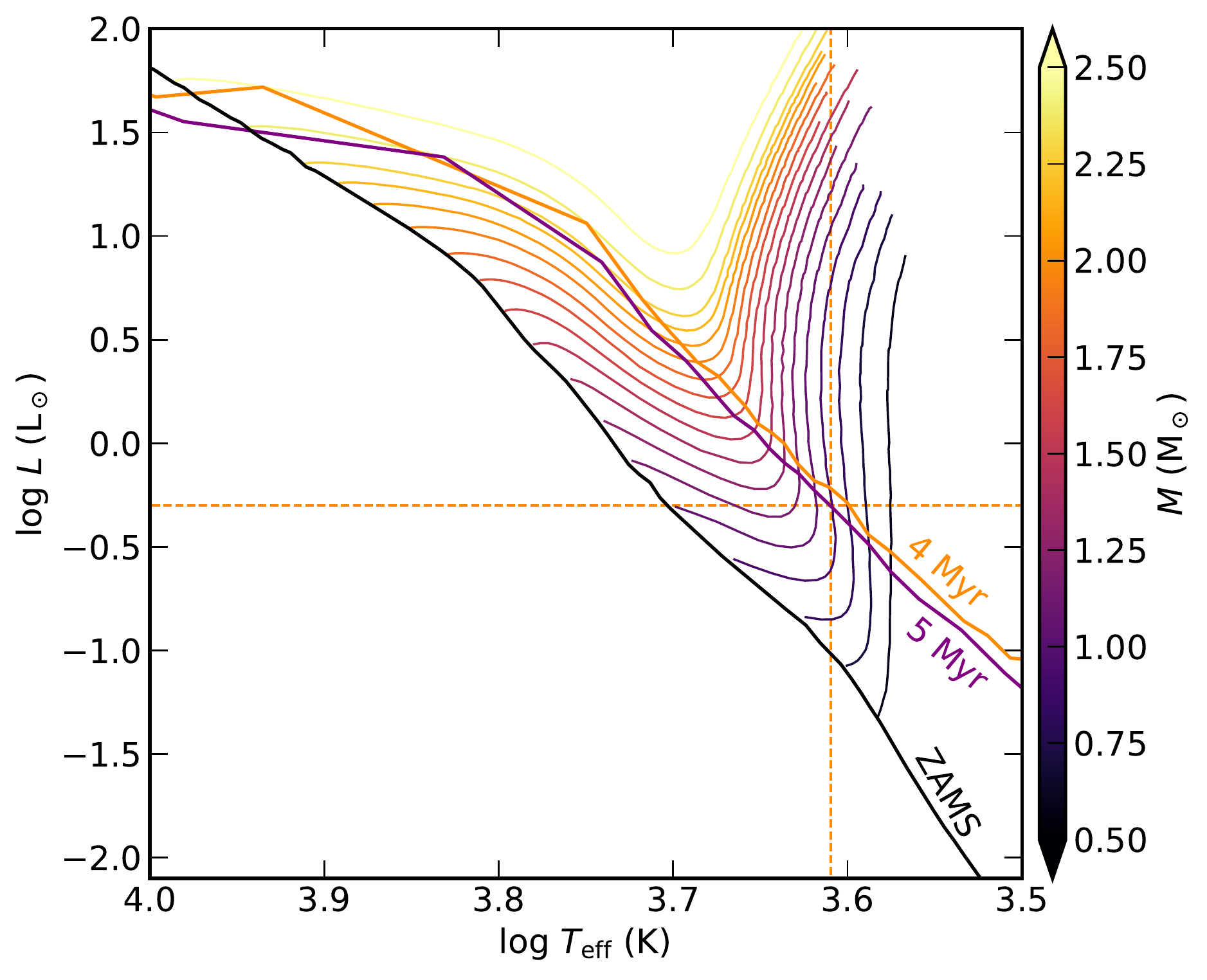}
    \caption{Hertzsprung-Russel diagram with stellar tracks from \citet{siess2000} for stellar masses from 0.5-2.5 M$_\odot$. The orange and purple lines denote the 4 and 5 Myr isochrones. The crosshair points at the effective temperature and luminosity, respectively, found in the MCMC runs, which agrees well with a $\sim$1 M$_\odot$ star derived from the dynamical fits to the CO emission.}
    \label{fig:HRdiagram}
\end{figure}

\subsection{Grain size distribution}
The dust grains in the atmospheric layers of the disk in HH~48~NE are constrained to have a size between 0.3 and 10 $\mu$m. 
The minimum grain size found in the disk is much larger than in the ISM dust distribution (typically 0.005 $\mu$m).
This result is consistent with previous findings of \citet{Pontoppidan2007}, showing that the high mid-infrared scattered light continuum in the Flying Saucer can only be explained by a dust distribution in the disk atmosphere with large grains (0.25 - 16~$\mu$m).
Our result is also consistent with previous simple comparisons of opacities with single grain sizes to Spitzer/IRS observations of the silicate feature at 10~$\mu$m. \citet{Olofsson2009} find that the observed silicate features in a large (96) sample of T Tauri disks are mainly produced by $\mu$m-sized grains. 
They show that there cannot be a large number of sub-micron-sized dust grains in the disk atmosphere, because their emission would overwhelm the few large grains.
Additionally, \citet{Tazaki2021} find evidence for >$\mu$m-sized grains in the upper layers of the HD 142527 disk, with an independent method, using resolved \ce{H2O} ice observations in coronagraphic imaging of scattered light \citep[see also][]{Honda2009}. 
The grain size limits are on the high end compared to the range found using the SEDs of 30 well-known (low inclination) protoplanetary disks \citep{kaeufer2023}.
\citet{kaeufer2023} used Bayesian analysis to find accurate parameter values with their uncertainties, and found a minimum grain size of typically 0.01-0.2 $\mu$m and a maximum grain size of typically 1-9 $\mu$m.

Grains larger than 1~$\mu$m are thought to settle quickly to the midplane \citep[<10$^5$ yr,][]{dullemond2004}, in the relevant regions (<100 au), due to the frictional force between the Keplerian orbit of the grain and the slightly sub-Keplerian orbit of the gas because of the pressure support.
Together with the very low levels of turbulence in lower disk regions \citep[e.g.,][and the references therein]{pinte2016,villenave2022,Flaherty2020}, this means that there has to be a vertical dust segregating mechanism that prevents the $\mu$m-sized dust grains from settling to the midplane or replenishes the $\mu$m-sized dust grains via grain-grain fragmentation, but does not affect the millimeter-sized dust grains observed with ALMA.
Some theoretical sources of turbulence can cause efficient vertical mixing, but do not reach all the way to the midplane (Zombie Vortex Instability, \citealt{Barranco2018}; see also \citealt{lesur2022}).
This particular source of turbulence is likely available in the strongly stratified regions of all protoplanetary disks, as long as there is modest grain settling and/or growth \citep{Barranco2018,lesur2022}. 
Only non-physically fast cooling and a high viscosity are known to suppress the instability \citep{Lesur2016}.
Mechanisms like these are a promising solution to the $\mu$m-sized dust grains in the upper layers of the disk, but need to be studied further in simulations to see if any direct observational signatures can serve as a proof of their existence.

In addition, there has to be a mechanism that efficiently removes sub-$\mu$m grains from the disk atmosphere, but leaves $\sim$$\mu$m sized dust grains in the upper disk. 
This could be possible if growth timescales for these small grains are significantly shorter than fragmentation timescales \citep[see e.g.,][]{Birnstiel2011,Windmark2012}.
This would require low levels of turbulence, or porous, fluffy grains that easily stick together to form larger grains \citep{Birnstiel2011}.
Since no ice is available in the disk atmospheres to make the grains stickier, due to the extreme temperatures and UV irradiation, this would require that the products of grain-grain fragmentation are inherently porous and fluffy.
An additional hypothesis is that the smallest dust grains can be lofted from the surface by disk winds, while the larger grains are too heavy to be carried away \citep{Olofsson2009,BoothR2021_dust_in_winds}.
Disk winds might be a ubiquitous driving mechanism for accretion in T Tauri protoplanetary disks \citep[see e.g.,][]{Tabone2022}, and may be launched from a substantial fraction of the disk surface \citep{Tabone2020}. 
Dust lofting by the wind is only efficient for smaller dust grains (<$\mu$m), and could significantly deplete the small dust population if sustained for $\sim$Myr timescales \citep{BoothR2021_dust_in_winds}.
Understanding the vertical dust segregation could be a key element of understanding the origin of turbulence in protoplanetary disks, and the weight-lifting capabilities of disk winds.

The exclusion of small nm-sized grains from the disk upper atmosphere has important implications for the disk chemistry. 
In many disk chemical models, the minimum grain size distribution is taken to be the same as for ISM grains ($\sim$5~nm) \citep[e.g.,][]{bruderer2012,Walsh2015}.
Decreasing the average surface area of dust grains in the disk atmosphere would naturally lead to a higher degree of UV radiation in the deeper disk layers.
An excess of UV radiation enhances photodissociation and photodesorption processes that ultimately result in a very rich chemistry \citep{Bergner2019_c2h_hcn,Bergner2021_Maps_HCN_CN,calahan_2022_uvchemistry}.

\subsection{Inner cavity}
We have shown that HH~48~NE likely has an inner cavity in the disk in both gas and dust of $\sim$50 au.
Similar cavities are found in a relatively large number of face-on disks (transition disks) \citep[$\sim$10\%;][]{ercolano2017}, but are so far not directly observed in edge-on protoplanetary disks. 
The radial structure of some of the disks in the large sample of \citet{Villenave2020} suggests that radial substructure is present in edge-on disks. 
However, detailed multi-wavelength modeling is necessary to exclude increased dust opacity due to high inclinations in combination with a high disk mass as an alternative explanation for disks with a flat radial flux distribution or central dip.
The inclusion of a cavity in our model is crucial to fit the SED in the mid-infrared.
The cavity allows photons emitted by warm dust in the inner-most region of the disk to escape if they have an ideal angle to scatter along our line of sight.
Similar conclusions are drawn from optical and near-infrared observations \citep[see][for a review]{Benisty2022}, and also lead to distinguishing Herbig Ae/Be stars in group I and group II sources. 
Group I being the sources with high levels of optical scattering due to large inner cavities, Group II being the sources without a cavity and much lower levels of scattering.

Without a cavity, we would need to add a 1400 K blackbody to the source spectrum, in a similar fashion to \citet{Pontoppidan2007}, to mimic increasing the amount of photons originating from the inner disk regions.
Adding a 1400 K blackbody manually solves the discrepancy in the SED fit, but is not self-consistent, as the 1400 K emission is thought to arise from the inner disk, not from the central star.
The difference in solid angle between a stellar point source with an additional 1400 K component and warm inner disk could result in crucial differences if one wants to model ice features in the mid-infrared, since the mid-infrared continuum would be dominated by photons scattering off of the disk that are produced in the warm inner disk, not in the star itself (see Appendix A in Paper II). 
We show that our model is self-consistent and naturally reproduces the flux in the mid-infrared.

Additional research, preferably with higher angular resolution ALMA  data, is necessary to see if the unnoticed existence of a cavity can resolve this issue in more sources. 
In edge-on disks, many of the smaller cavities or rings will be hidden behind the optically thick dust in the midplane, and multi-wavelength analysis is necessary to constrain it. 
In the specific case of the Flying Saucer, where a similar deficit in mid-infrared flux is found as in our model without a cavity \citep{Pontoppidan2007}, the continuum hints at a double-peaked continuum in the radial direction \citep{guilloteau2016}, even though the dip is weaker and less resolved than in HH~48.
Additional detailed multi-wavelength studies of other edge-on disks are necessary to see if more of them harbor a cavity. 

\section{Conclusions}
\label{sec:conclusion}
We have developed a disk model using the radiative transfer code \texttt{RADMC-3D}  that includes anisotropic scattering.
We have run MCMC fits to HST observations, ALMA observations and the SED for HH~48~NE to constrain the geometry and physical features of this edge-on Class II system. We can conclude the following:

\begin{itemize}
\item The best fitting model (SED run) reproduces both the spatially resolved features of the emission in scattered light observations, mm dust continuum observations, CO gas observations, and the SED.
\item HH~48~NE is a K7 star with a mass of 1-1.4 M$_\odot$, an effective temperature of $\sim$4150 K and a luminosity of $\sim$0.4 L$_\odot$. 
These values agree with the estimated age of the Chamaeleon I star-forming region following evolutionary stellar tracks from the literature. 
\item HH~48~NE is a typical T Tauri disk in many ways, but stands out due to the lack of small (<0.3~$\mu$m) grains in the disk atmosphere. 
Mechanisms that remove the small dust grains can be the result of efficient sticking of the smallest grains, or lofting in large scale disk winds.
The exclusion of small grains may have a major impact on the chemistry in the disk, as UV radiation can likely penetrate much deeper into the disk due to the decreased average surface area of the grains.
\item The disk in HH~48~NE likely has a $\sim$50 au inner cavity in gas and dust that is mostly invisible due to optically thick dust. 
The cavity boosts the mid-infrared emission in the SED and allows for a unique model that is consistent with the scattered light observations and the mm continuum observations at the same time.
\end{itemize}

Understanding the mechanisms that set grain size distributions and physical properties of protoplanetary disks is crucial. Edge-on disks can be a unique tool for constraining some parameters that are hard to constrain in less inclined disks, such as dust settling and the grain size distribution.

\begin{acknowledgements}
We would like to thank the anonymous referee for suggestions that improved the manuscript.
We also thank Ewine van Dishoeck for useful discussions and constructive comments on the manuscript.
Astrochemistry in Leiden is supported by the Netherlands Research School for Astronomy (NOVA), by funding from the European Research Council (ERC) under the European Union’s Horizon 2020 research and innovation programme (grant agreement No. 101019751 MOLDISK).
M.N.D. acknowledges the Swiss National Science Foundation (SNSF) Ambizione grant no. 180079, the Center for Space and Habitability (CSH) Fellowship, and the IAU Gruber Foundation Fellowship.

This paper makes use of the following ALMA data: ADS/JAO.ALMA\#2016.1.00460.S, ADS/JAO.ALMA\#2019.1.01792.S. ALMA is a partnership of ESO (representing its member states), NSF (USA) and NINS (Japan), together with NRC (Canada), MOST and ASIAA (Taiwan), and KASI (Republic of Korea), in cooperation with the Republic of Chile. The Joint ALMA Observatory is operated by ESO, AUI/NRAO and NAOJ. The National Radio Astronomy Observatory is a facility of the National Science Foundation operated under cooperative agreement by Associated Universities, Inc.

This work makes use of the following software: \texttt{numpy} \citep{numpy}, \texttt{matplotlib} \citep{matplotlib}, \texttt{astropy} \citep{astropy2013,astropy2018}, \texttt{bettermoments} \citep{Teague18}, CASA \citep{McMullin07}, \texttt{eddy} \citep{Teague19_eddy}, \texttt{emcee} \citep{Foreman_Mackey13}, \texttt{OpTool} \citep{Dominik2021optool}, \texttt{RADMC-3D} \citep{Dullemond2012radmc3d}.
\end{acknowledgements}

\bibliographystyle{aa}
\bibliography{refs.bib}

\begin{thebibliography}{81}
\expandafter\ifx\csname natexlab\endcsname\relax\def\natexlab#1{#1}\fi

\bibitem[{{Andrews} {et~al.}(2011){Andrews}, {Wilner}, {Espaillat}, {Hughes},
  {Dullemond}, {McClure}, {Qi}, \& {Brown}}]{andrews2011}
{Andrews}, S.~M., {Wilner}, D.~J., {Espaillat}, C., {et~al.} 2011, \apj, 732,
  42

\bibitem[{{Angelo} {et~al.}(2023){Angelo}, {Duch{\^e}ne}, {Stapelfeldt},
  {Telkamp}, {M{\'e}nard}, {Padgett}, {van der Plas}, {Villenave}, {Pinte},
  {Wolff}, {Fischer}, \& {Perrin}}]{Angelo2023}
{Angelo}, I., {Duch{\^e}ne}, G., {Stapelfeldt}, K., {et~al.} 2023, arXiv
  e-prints, arXiv:2302.04891

\bibitem[{{Ansdell} {et~al.}(2016){Ansdell}, {Williams}, {van der Marel},
  {Carpenter}, {Guidi}, {Hogerheijde}, {Mathews}, {Manara}, {Miotello},
  {Natta}, {Oliveira}, {Tazzari}, {Testi}, {van Dishoeck}, \& {van
  Terwisga}}]{Ansdell2016}
{Ansdell}, M., {Williams}, J.~P., {van der Marel}, N., {et~al.} 2016, \apj,
  828, 46

\bibitem[{{Astropy Collaboration} {et~al.}(2018){Astropy Collaboration},
  {Price-Whelan}, {Sip{\H{o}}cz}, {G{\"u}nther}, {Lim}, {Crawford}, {Conseil},
  {Shupe}, {Craig}, {Dencheva}, {Ginsburg}, {VanderPlas}, {Bradley},
  {P{\'e}rez-Su{\'a}rez}, {de Val-Borro}, {Aldcroft}, {Cruz}, {Robitaille},
  {Tollerud}, {Ardelean}, {Babej}, {Bach}, {Bachetti}, {Bakanov}, {Bamford},
  {Barentsen}, {Barmby}, {Baumbach}, {Berry}, {Biscani}, {Boquien}, {Bostroem},
  {Bouma}, {Brammer}, {Bray}, {Breytenbach}, {Buddelmeijer}, {Burke},
  {Calderone}, {Cano Rodr{\'\i}guez}, {Cara}, {Cardoso}, {Cheedella}, {Copin},
  {Corrales}, {Crichton}, {D'Avella}, {Deil}, {Depagne}, {Dietrich}, {Donath},
  {Droettboom}, {Earl}, {Erben}, {Fabbro}, {Ferreira}, {Finethy}, {Fox},
  {Garrison}, {Gibbons}, {Goldstein}, {Gommers}, {Greco}, {Greenfield},
  {Groener}, {Grollier}, {Hagen}, {Hirst}, {Homeier}, {Horton}, {Hosseinzadeh},
  {Hu}, {Hunkeler}, {Ivezi{\'c}}, {Jain}, {Jenness}, {Kanarek}, {Kendrew},
  {Kern}, {Kerzendorf}, {Khvalko}, {King}, {Kirkby}, {Kulkarni}, {Kumar},
  {Lee}, {Lenz}, {Littlefair}, {Ma}, {Macleod}, {Mastropietro}, {McCully},
  {Montagnac}, {Morris}, {Mueller}, {Mumford}, {Muna}, {Murphy}, {Nelson},
  {Nguyen}, {Ninan}, {N{\"o}the}, {Ogaz}, {Oh}, {Parejko}, {Parley}, {Pascual},
  {Patil}, {Patil}, {Plunkett}, {Prochaska}, {Rastogi}, {Reddy Janga},
  {Sabater}, {Sakurikar}, {Seifert}, {Sherbert}, {Sherwood-Taylor}, {Shih},
  {Sick}, {Silbiger}, {Singanamalla}, {Singer}, {Sladen}, {Sooley},
  {Sornarajah}, {Streicher}, {Teuben}, {Thomas}, {Tremblay}, {Turner},
  {Terr{\'o}n}, {van Kerkwijk}, {de la Vega}, {Watkins}, {Weaver}, {Whitmore},
  {Woillez}, {Zabalza}, \& {Astropy Contributors}}]{astropy2018}
{Astropy Collaboration}, {Price-Whelan}, A.~M., {Sip{\H{o}}cz}, B.~M., {et~al.}
  2018, \aj, 156, 123

\bibitem[{{Astropy Collaboration} {et~al.}(2013){Astropy Collaboration},
  {Robitaille}, {Tollerud}, {Greenfield}, {Droettboom}, {Bray}, {Aldcroft},
  {Davis}, {Ginsburg}, {Price-Whelan}, {Kerzendorf}, {Conley}, {Crighton},
  {Barbary}, {Muna}, {Ferguson}, {Grollier}, {Parikh}, {Nair}, {Unther},
  {Deil}, {Woillez}, {Conseil}, {Kramer}, {Turner}, {Singer}, {Fox}, {Weaver},
  {Zabalza}, {Edwards}, {Azalee Bostroem}, {Burke}, {Casey}, {Crawford},
  {Dencheva}, {Ely}, {Jenness}, {Labrie}, {Lim}, {Pierfederici}, {Pontzen},
  {Ptak}, {Refsdal}, {Servillat}, \& {Streicher}}]{astropy2013}
{Astropy Collaboration}, {Robitaille}, T.~P., {Tollerud}, E.~J., {et~al.} 2013,
  \aap, 558, A33

\bibitem[{{Barranco} {et~al.}(2018){Barranco}, {Pei}, \&
  {Marcus}}]{Barranco2018}
{Barranco}, J.~A., {Pei}, S., \& {Marcus}, P.~S. 2018, \apj, 869, 127

\bibitem[{{Benisty} {et~al.}(2022){Benisty}, {Dominik}, {Follette}, {Garufi},
  {Ginski}, {Hashimoto}, {Keppler}, {Kley}, \& {Monnier}}]{Benisty2022}
{Benisty}, M., {Dominik}, C., {Follette}, K., {et~al.} 2022, arXiv e-prints,
  arXiv:2203.09991

\bibitem[{{Bergner} {et~al.}(2019){Bergner}, {{\"O}berg}, {Bergin}, {Loomis},
  {Pegues}, \& {Qi}}]{Bergner2019_c2h_hcn}
{Bergner}, J.~B., {{\"O}berg}, K.~I., {Bergin}, E.~A., {et~al.} 2019, \apj,
  876, 25

\bibitem[{{Bergner} {et~al.}(2021){Bergner}, {{\"O}berg}, {Guzm{\'a}n}, {Law},
  {Loomis}, {Cataldi}, {Bosman}, {Aikawa}, {Andrews}, {Bergin}, {Booth},
  {Cleeves}, {Czekala}, {Huang}, {Ilee}, {Le Gal}, {Long}, {Nomura},
  {M{\'e}nard}, {Qi}, {Schwarz}, {Teague}, {Tsukagoshi}, {Walsh}, {Wilner}, \&
  {Yamato}}]{Bergner2021_Maps_HCN_CN}
{Bergner}, J.~B., {{\"O}berg}, K.~I., {Guzm{\'a}n}, V.~V., {et~al.} 2021,
  \apjs, 257, 11

\bibitem[{{Birnstiel} {et~al.}(2011){Birnstiel}, {Ormel}, \&
  {Dullemond}}]{Birnstiel2011}
{Birnstiel}, T., {Ormel}, C.~W., \& {Dullemond}, C.~P. 2011, \aap, 525, A11

\bibitem[{{Booth} \& {Clarke}(2021)}]{BoothR2021_dust_in_winds}
{Booth}, R.~A. \& {Clarke}, C.~J. 2021, \mnras, 502, 1569

\bibitem[{{Bruderer}(2013)}]{bruderer2013}
{Bruderer}, S. 2013, \aap, 559, A46

\bibitem[{{Bruderer} {et~al.}(2012){Bruderer}, {van Dishoeck}, {Doty}, \&
  {Herczeg}}]{bruderer2012}
{Bruderer}, S., {van Dishoeck}, E.~F., {Doty}, S.~D., \& {Herczeg}, G.~J. 2012,
  \aap, 541, A91

\bibitem[{{Calahan} {et~al.}(2023){Calahan}, {Bergin}, {Bosman}, {Rich},
  {Andrews}, {Bergner}, {Cleeves}, {Guzm{\'a}n}, {Huang}, {Ilee}, {Law}, {Le
  Gal}, {{\"O}berg}, {Teague}, {Walsh}, {Wilner}, \&
  {Zhang}}]{calahan_2022_uvchemistry}
{Calahan}, J.~K., {Bergin}, E.~A., {Bosman}, A.~D., {et~al.} 2023, Nature
  Astronomy, 7, 49

\bibitem[{{Cardelli} {et~al.}(1989){Cardelli}, {Clayton}, \&
  {Mathis}}]{1989ApJ...345..245C}
{Cardelli}, J.~A., {Clayton}, G.~C., \& {Mathis}, J.~S. 1989, \apj, 345, 245

\bibitem[{{Dominik} {et~al.}(2021){Dominik}, {Min}, \&
  {Tazaki}}]{Dominik2021optool}
{Dominik}, C., {Min}, M., \& {Tazaki}, R. 2021, {OpTool: Command-line driven
  tool for creating complex dust opacities}, Astrophysics Source Code Library,
  record ascl:2104.010

\bibitem[{{Dullemond} \& {Dominik}(2004)}]{dullemond2004}
{Dullemond}, C.~P. \& {Dominik}, C. 2004, \aap, 421, 1075

\bibitem[{{Dullemond} {et~al.}(2012){Dullemond}, {Juhasz}, {Pohl}, {Sereshti},
  {Shetty}, {Peters}, {Commercon}, \& {Flock}}]{Dullemond2012radmc3d}
{Dullemond}, C.~P., {Juhasz}, A., {Pohl}, A., {et~al.} 2012, {RADMC-3D: A
  multi-purpose radiative transfer tool}, Astrophysics Source Code Library,
  record ascl:1202.015

\bibitem[{{Dunham} {et~al.}(2016){Dunham}, {Offner}, {Pineda}, {Bourke},
  {Tobin}, {Arce}, {Chen}, {Di Francesco}, {Johnstone}, {Lee}, {Myers},
  {Price}, {Sadavoy}, \& {Schnee}}]{Dunham2016}
{Dunham}, M.~M., {Offner}, S. S.~R., {Pineda}, J.~E., {et~al.} 2016, \apj, 823,
  160

\bibitem[{{Dutrey} {et~al.}(2017){Dutrey}, {Guilloteau}, {Pi{\'e}tu},
  {Chapillon}, {Wakelam}, {Di Folco}, {Stoecklin}, {Denis-Alpizar}, {Gorti},
  {Teague}, {Henning}, {Semenov}, \& {Grosso}}]{Dutrey2017}
{Dutrey}, A., {Guilloteau}, S., {Pi{\'e}tu}, V., {et~al.} 2017, \aap, 607, A130

\bibitem[{{Earl} \& {Deem}(2005)}]{Earl2005PTmcmc}
{Earl}, D.~J. \& {Deem}, M.~W. 2005, Physical Chemistry Chemical Physics
  (Incorporating Faraday Transactions), 7, 3910

\bibitem[{{Ercolano} \& {Pascucci}(2017)}]{ercolano2017}
{Ercolano}, B. \& {Pascucci}, I. 2017, Royal Society Open Science, 4, 170114

\bibitem[{{Espaillat} {et~al.}(2019){Espaillat}, {Mac{\'\i}as},
  {Hern{\'a}ndez}, \& {Robinson}}]{espaillat2019}
{Espaillat}, C.~C., {Mac{\'\i}as}, E., {Hern{\'a}ndez}, J., \& {Robinson}, C.
  2019, \apjl, 877, L34

\bibitem[{{Flaherty} {et~al.}(2020){Flaherty}, {Hughes}, {Simon}, {Qi}, {Bai},
  {Bulatek}, {Andrews}, {Wilner}, \& {K{\'o}sp{\'a}l}}]{Flaherty2020}
{Flaherty}, K., {Hughes}, A.~M., {Simon}, J.~B., {et~al.} 2020, \apj, 895, 109

\bibitem[{{Flores} {et~al.}(2021){Flores}, {Duch{\^e}ne}, {Wolff}, {Villenave},
  {Stapelfeldt}, {Williams}, {Pinte}, {Padgett}, {Connelley}, {van der Plas},
  {M{\'e}nard}, \& {Perrin}}]{Flores2021}
{Flores}, C., {Duch{\^e}ne}, G., {Wolff}, S., {et~al.} 2021, \aj, 161, 239

\bibitem[{{Foreman-Mackey} {et~al.}(2013){Foreman-Mackey}, {Hogg}, {Lang}, \&
  {Goodman}}]{Foreman_Mackey13}
{Foreman-Mackey}, D., {Hogg}, D.~W., {Lang}, D., \& {Goodman}, J. 2013, \pasp,
  125, 306

\bibitem[{{Furlan} {et~al.}(2011){Furlan}, {Luhman}, {Espaillat}, {D'Alessio},
  {Adame}, {Manoj}, {Kim}, {Watson}, {Forrest}, {McClure}, {Calvet}, {Sargent},
  {Green}, \& {Fischer}}]{Furlan2011}
{Furlan}, E., {Luhman}, K.~L., {Espaillat}, C., {et~al.} 2011, \apjs, 195, 3

\bibitem[{{Gaia Collaboration} {et~al.}(2021){Gaia Collaboration}, {Brown},
  {Vallenari}, {Prusti}, {de Bruijne}, {Babusiaux}, {Biermann}, {Creevey},
  {Evans}, {Eyer}, {Hutton}, {Jansen}, {Jordi}, {Klioner}, {Lammers},
  {Lindegren}, {Luri}, {Mignard}, {Panem}, {Pourbaix}, {Randich}, {Sartoretti},
  {Soubiran}, {Walton}, {Arenou}, {Bailer-Jones}, {Bastian}, {Cropper},
  {Drimmel}, {Katz}, {Lattanzi}, {van Leeuwen}, {Bakker}, {Cacciari},
  {Casta{\~n}eda}, {De Angeli}, {Ducourant}, {Fabricius}, {Fouesneau},
  {Fr{\'e}mat}, {Guerra}, {Guerrier}, {Guiraud}, {Jean-Antoine Piccolo},
  {Masana}, {Messineo}, {Mowlavi}, {Nicolas}, {Nienartowicz}, {Pailler},
  {Panuzzo}, {Riclet}, {Roux}, {Seabroke}, {Sordo}, {Tanga}, {Th{\'e}venin},
  {Gracia-Abril}, {Portell}, {Teyssier}, {Altmann}, {Andrae}, {Bellas-Velidis},
  {Benson}, {Berthier}, {Blomme}, {Brugaletta}, {Burgess}, {Busso}, {Carry},
  {Cellino}, {Cheek}, {Clementini}, {Damerdji}, {Davidson}, {Delchambre},
  {Dell'Oro}, {Fern{\'a}ndez-Hern{\'a}ndez}, {Galluccio}, {Garc{\'\i}a-Lario},
  {Garcia-Reinaldos}, {Gonz{\'a}lez-N{\'u}{\~n}ez}, {Gosset}, {Haigron},
  {Halbwachs}, {Hambly}, {Harrison}, {Hatzidimitriou}, {Heiter},
  {Hern{\'a}ndez}, {Hestroffer}, {Hodgkin}, {Holl}, {Jan{\ss}en}, {Jevardat de
  Fombelle}, {Jordan}, {Krone-Martins}, {Lanzafame}, {L{\"o}ffler}, {Lorca},
  {Manteiga}, {Marchal}, {Marrese}, {Moitinho}, {Mora}, {Muinonen}, {Osborne},
  {Pancino}, {Pauwels}, {Petit}, {Recio-Blanco}, {Richards}, {Riello},
  {Rimoldini}, {Robin}, {Roegiers}, {Rybizki}, {Sarro}, {Siopis}, {Smith},
  {Sozzetti}, {Ulla}, {Utrilla}, {van Leeuwen}, {van Reeven}, {Abbas}, {Abreu
  Aramburu}, {Accart}, {Aerts}, {Aguado}, {Ajaj}, {Altavilla}, {{\'A}lvarez},
  {{\'A}lvarez Cid-Fuentes}, {Alves}, {Anderson}, {Anglada Varela}, {Antoja},
  {Audard}, {Baines}, {Baker}, {Balaguer-N{\'u}{\~n}ez}, {Balbinot}, {Balog},
  {Barache}, {Barbato}, {Barros}, {Barstow}, {Bartolom{\'e}}, {Bassilana},
  {Bauchet}, {Baudesson-Stella}, {Becciani}, {Bellazzini}, {Bernet}, {Bertone},
  {Bianchi}, {Blanco-Cuaresma}, {Boch}, {Bombrun}, {Bossini}, {Bouquillon},
  {Bragaglia}, {Bramante}, {Breedt}, {Bressan}, {Brouillet}, {Bucciarelli},
  {Burlacu}, {Busonero}, {Butkevich}, {Buzzi}, {Caffau}, {Cancelliere},
  {C{\'a}novas}, {Cantat-Gaudin}, {Carballo}, {Carlucci}, {Carnerero},
  {Carrasco}, {Casamiquela}, {Castellani}, {Castro-Ginard}, {Castro Sampol},
  {Chaoul}, {Charlot}, {Chemin}, {Chiavassa}, {Cioni}, {Comoretto}, {Cooper},
  {Cornez}, {Cowell}, {Crifo}, {Crosta}, {Crowley}, {Dafonte}, {Dapergolas},
  {David}, {David}, {de Laverny}, {De Luise}, {De March}, {De Ridder}, {de
  Souza}, {de Teodoro}, {de Torres}, {del Peloso}, {del Pozo}, {Delbo},
  {Delgado}, {Delgado}, {Delisle}, {Di Matteo}, {Diakite}, {Diener},
  {Distefano}, {Dolding}, {Eappachen}, {Edvardsson}, {Enke}, {Esquej}, {Fabre},
  {Fabrizio}, {Faigler}, {Fedorets}, {Fernique}, {Fienga}, {Figueras},
  {Fouron}, {Fragkoudi}, {Fraile}, {Franke}, {Gai}, {Garabato},
  {Garcia-Gutierrez}, {Garc{\'\i}a-Torres}, {Garofalo}, {Gavras}, {Gerlach},
  {Geyer}, {Giacobbe}, {Gilmore}, {Girona}, {Giuffrida}, {Gomel}, {Gomez},
  {Gonzalez-Santamaria}, {Gonz{\'a}lez-Vidal}, {Granvik},
  {Guti{\'e}rrez-S{\'a}nchez}, {Guy}, {Hauser}, {Haywood}, {Helmi}, {Hidalgo},
  {Hilger}, {H{\l}adczuk}, {Hobbs}, {Holland}, {Huckle}, {Jasniewicz},
  {Jonker}, {Juaristi Campillo}, {Julbe}, {Karbevska}, {Kervella}, {Khanna},
  {Kochoska}, {Kontizas}, {Kordopatis}, {Korn}, {Kostrzewa-Rutkowska},
  {Kruszy{\'n}ska}, {Lambert}, {Lanza}, {Lasne}, {Le Campion}, {Le Fustec},
  {Lebreton}, {Lebzelter}, {Leccia}, {Leclerc}, {Lecoeur-Taibi}, {Liao},
  {Licata}, {Lindstr{\o}m}, {Lister}, {Livanou}, {Lobel}, {Madrero Pardo},
  {Managau}, {Mann}, {Marchant}, {Marconi}, {Marcos Santos}, {Marinoni},
  {Marocco}, {Marshall}, {Martin Polo}, {Mart{\'\i}n-Fleitas}, {Masip},
  {Massari}, {Mastrobuono-Battisti}, {Mazeh}, {McMillan}, {Messina},
  {Michalik}, {Millar}, {Mints}, {Molina}, {Molinaro}, {Moln{\'a}r},
  {Montegriffo}, {Mor}, {Morbidelli}, {Morel}, {Morris}, {Mulone}, {Munoz},
  {Muraveva}, {Murphy}, {Musella}, {Noval}, {Ord{\'e}novic}, {Orr{\`u}},
  {Osinde}, {Pagani}, {Pagano}, {Palaversa}, {Palicio}, {Panahi}, {Pawlak},
  {Pe{\~n}alosa Esteller}, {Penttil{\"a}}, {Piersimoni}, {Pineau}, {Plachy},
  {Plum}, {Poggio}, {Poretti}, {Poujoulet}, {Pr{\v{s}}a}, {Pulone}, {Racero},
  {Ragaini}, {Rainer}, {Raiteri}, {Rambaux}, {Ramos}, {Ramos-Lerate}, {Re
  Fiorentin}, {Regibo}, {Reyl{\'e}}, {Ripepi}, {Riva}, {Rixon}, {Robichon},
  {Robin}, {Roelens}, {Rohrbasser}, {Romero-G{\'o}mez}, {Rowell}, {Royer},
  {Rybicki}, {Sadowski}, {Sagrist{\`a} Sell{\'e}s}, {Sahlmann}, {Salgado},
  {Salguero}, {Samaras}, {Sanchez Gimenez}, {Sanna}, {Santove{\~n}a},
  {Sarasso}, {Schultheis}, {Sciacca}, {Segol}, {Segovia}, {S{\'e}gransan},
  {Semeux}, {Shahaf}, {Siddiqui}, {Siebert}, {Siltala}, {Slezak}, {Smart},
  {Solano}, {Solitro}, {Souami}, {Souchay}, {Spagna}, {Spoto}, {Steele},
  {Steidelm{\"u}ller}, {Stephenson}, {S{\"u}veges}, {Szabados}, {Szegedi-Elek},
  {Taris}, {Tauran}, {Taylor}, {Teixeira}, {Thuillot}, {Tonello}, {Torra},
  {Torra}, {Turon}, {Unger}, {Vaillant}, {van Dillen}, {Vanel}, {Vecchiato},
  {Viala}, {Vicente}, {Voutsinas}, {Weiler}, {Wevers}, {Wyrzykowski}, {Yoldas},
  {Yvard}, {Zhao}, {Zorec}, {Zucker}, {Zurbach}, \& {Zwitter}}]{GAIADR3}
{Gaia Collaboration}, {Brown}, A.~G.~A., {Vallenari}, A., {et~al.} 2021, \aap,
  649, A1

\bibitem[{{Gaia Collaboration} {et~al.}(2016){Gaia Collaboration}, {Prusti},
  {de Bruijne}, {Brown}, {Vallenari}, {Babusiaux}, {Bailer-Jones}, {Bastian},
  {Biermann}, {Evans}, {Eyer}, {Jansen}, {Jordi}, {Klioner}, {Lammers},
  {Lindegren}, {Luri}, {Mignard}, {Milligan}, {Panem}, {Poinsignon},
  {Pourbaix}, {Randich}, {Sarri}, {Sartoretti}, {Siddiqui}, {Soubiran},
  {Valette}, {van Leeuwen}, {Walton}, {Aerts}, {Arenou}, {Cropper}, {Drimmel},
  {H{\o}g}, {Katz}, {Lattanzi}, {O'Mullane}, {Grebel}, {Holland}, {Huc},
  {Passot}, {Bramante}, {Cacciari}, {Casta{\~n}eda}, {Chaoul}, {Cheek}, {De
  Angeli}, {Fabricius}, {Guerra}, {Hern{\'a}ndez}, {Jean-Antoine-Piccolo},
  {Masana}, {Messineo}, {Mowlavi}, {Nienartowicz}, {Ord{\'o}{\~n}ez-Blanco},
  {Panuzzo}, {Portell}, {Richards}, {Riello}, {Seabroke}, {Tanga},
  {Th{\'e}venin}, {Torra}, {Els}, {Gracia-Abril}, {Comoretto},
  {Garcia-Reinaldos}, {Lock}, {Mercier}, {Altmann}, {Andrae}, {Astraatmadja},
  {Bellas-Velidis}, {Benson}, {Berthier}, {Blomme}, {Busso}, {Carry},
  {Cellino}, {Clementini}, {Cowell}, {Creevey}, {Cuypers}, {Davidson}, {De
  Ridder}, {de Torres}, {Delchambre}, {Dell'Oro}, {Ducourant}, {Fr{\'e}mat},
  {Garc{\'\i}a-Torres}, {Gosset}, {Halbwachs}, {Hambly}, {Harrison}, {Hauser},
  {Hestroffer}, {Hodgkin}, {Huckle}, {Hutton}, {Jasniewicz}, {Jordan},
  {Kontizas}, {Korn}, {Lanzafame}, {Manteiga}, {Moitinho}, {Muinonen},
  {Osinde}, {Pancino}, {Pauwels}, {Petit}, {Recio-Blanco}, {Robin}, {Sarro},
  {Siopis}, {Smith}, {Smith}, {Sozzetti}, {Thuillot}, {van Reeven}, {Viala},
  {Abbas}, {Abreu Aramburu}, {Accart}, {Aguado}, {Allan}, {Allasia},
  {Altavilla}, {{\'A}lvarez}, {Alves}, {Anderson}, {Andrei}, {Anglada Varela},
  {Antiche}, {Antoja}, {Ant{\'o}n}, {Arcay}, {Atzei}, {Ayache}, {Bach},
  {Baker}, {Balaguer-N{\'u}{\~n}ez}, {Barache}, {Barata}, {Barbier}, {Barblan},
  {Baroni}, {Barrado y Navascu{\'e}s}, {Barros}, {Barstow}, {Becciani},
  {Bellazzini}, {Bellei}, {Bello Garc{\'\i}a}, {Belokurov}, {Bendjoya},
  {Berihuete}, {Bianchi}, {Bienaym{\'e}}, {Billebaud}, {Blagorodnova},
  {Blanco-Cuaresma}, {Boch}, {Bombrun}, {Borrachero}, {Bouquillon}, {Bourda},
  {Bouy}, {Bragaglia}, {Breddels}, {Brouillet}, {Br{\"u}semeister},
  {Bucciarelli}, {Budnik}, {Burgess}, {Burgon}, {Burlacu}, {Busonero}, {Buzzi},
  {Caffau}, {Cambras}, {Campbell}, {Cancelliere}, {Cantat-Gaudin}, {Carlucci},
  {Carrasco}, {Castellani}, {Charlot}, {Charnas}, {Charvet}, {Chassat},
  {Chiavassa}, {Clotet}, {Cocozza}, {Collins}, {Collins}, {Costigan}, {Crifo},
  {Cross}, {Crosta}, {Crowley}, {Dafonte}, {Damerdji}, {Dapergolas}, {David},
  {David}, {De Cat}, {de Felice}, {de Laverny}, {De Luise}, {De March}, {de
  Martino}, {de Souza}, {Debosscher}, {del Pozo}, {Delbo}, {Delgado},
  {Delgado}, {di Marco}, {Di Matteo}, {Diakite}, {Distefano}, {Dolding}, {Dos
  Anjos}, {Drazinos}, {Dur{\'a}n}, {Dzigan}, {Ecale}, {Edvardsson}, {Enke},
  {Erdmann}, {Escolar}, {Espina}, {Evans}, {Eynard Bontemps}, {Fabre},
  {Fabrizio}, {Faigler}, {Falc{\~a}o}, {Farr{\`a}s Casas}, {Faye}, {Federici},
  {Fedorets}, {Fern{\'a}ndez-Hern{\'a}ndez}, {Fernique}, {Fienga}, {Figueras},
  {Filippi}, {Findeisen}, {Fonti}, {Fouesneau}, {Fraile}, {Fraser}, {Fuchs},
  {Furnell}, {Gai}, {Galleti}, {Galluccio}, {Garabato}, {Garc{\'\i}a-Sedano},
  {Gar{\'e}}, {Garofalo}, {Garralda}, {Gavras}, {Gerssen}, {Geyer}, {Gilmore},
  {Girona}, {Giuffrida}, {Gomes}, {Gonz{\'a}lez-Marcos},
  {Gonz{\'a}lez-N{\'u}{\~n}ez}, {Gonz{\'a}lez-Vidal}, {Granvik}, {Guerrier},
  {Guillout}, {Guiraud}, {G{\'u}rpide}, {Guti{\'e}rrez-S{\'a}nchez}, {Guy},
  {Haigron}, {Hatzidimitriou}, {Haywood}, {Heiter}, {Helmi}, {Hobbs},
  {Hofmann}, {Holl}, {Holland}, {Hunt}, {Hypki}, {Icardi}, {Irwin}, {Jevardat
  de Fombelle}, {Jofr{\'e}}, {Jonker}, {Jorissen}, {Julbe}, {Karampelas},
  {Kochoska}, {Kohley}, {Kolenberg}, {Kontizas}, {Koposov}, {Kordopatis},
  {Koubsky}, {Kowalczyk}, {Krone-Martins}, {Kudryashova}, {Kull}, {Bachchan},
  {Lacoste-Seris}, {Lanza}, {Lavigne}, {Le Poncin-Lafitte}, {Lebreton},
  {Lebzelter}, {Leccia}, {Leclerc}, {Lecoeur-Taibi}, {Lemaitre}, {Lenhardt},
  {Leroux}, {Liao}, {Licata}, {Lindstr{\o}m}, {Lister}, {Livanou}, {Lobel},
  {L{\"o}ffler}, {L{\'o}pez}, {Lopez-Lozano}, {Lorenz}, {Loureiro},
  {MacDonald}, {Magalh{\~a}es Fernandes}, {Managau}, {Mann}, {Mantelet},
  {Marchal}, {Marchant}, {Marconi}, {Marie}, {Marinoni}, {Marrese},
  {Marschalk{\'o}}, {Marshall}, {Mart{\'\i}n-Fleitas}, {Martino}, {Mary},
  {Matijevi{\v{c}}}, {Mazeh}, {McMillan}, {Messina}, {Mestre}, {Michalik},
  {Millar}, {Miranda}, {Molina}, {Molinaro}, {Molinaro}, {Moln{\'a}r},
  {Moniez}, {Montegriffo}, {Monteiro}, {Mor}, {Mora}, {Morbidelli}, {Morel},
  {Morgenthaler}, {Morley}, {Morris}, {Mulone}, {Muraveva}, {Musella},
  {Narbonne}, {Nelemans}, {Nicastro}, {Noval}, {Ord{\'e}novic},
  {Ordieres-Mer{\'e}}, {Osborne}, {Pagani}, {Pagano}, {Pailler}, {Palacin},
  {Palaversa}, {Parsons}, {Paulsen}, {Pecoraro}, {Pedrosa}, {Pentik{\"a}inen},
  {Pereira}, {Pichon}, {Piersimoni}, {Pineau}, {Plachy}, {Plum}, {Poujoulet},
  {Pr{\v{s}}a}, {Pulone}, {Ragaini}, {Rago}, {Rambaux}, {Ramos-Lerate},
  {Ranalli}, {Rauw}, {Read}, {Regibo}, {Renk}, {Reyl{\'e}}, {Ribeiro},
  {Rimoldini}, {Ripepi}, {Riva}, {Rixon}, {Roelens}, {Romero-G{\'o}mez},
  {Rowell}, {Royer}, {Rudolph}, {Ruiz-Dern}, {Sadowski}, {Sagrist{\`a}
  Sell{\'e}s}, {Sahlmann}, {Salgado}, {Salguero}, {Sarasso}, {Savietto},
  {Schnorhk}, {Schultheis}, {Sciacca}, {Segol}, {Segovia}, {Segransan},
  {Serpell}, {Shih}, {Smareglia}, {Smart}, {Smith}, {Solano}, {Solitro},
  {Sordo}, {Soria Nieto}, {Souchay}, {Spagna}, {Spoto}, {Stampa}, {Steele},
  {Steidelm{\"u}ller}, {Stephenson}, {Stoev}, {Suess}, {S{\"u}veges}, {Surdej},
  {Szabados}, {Szegedi-Elek}, {Tapiador}, {Taris}, {Tauran}, {Taylor},
  {Teixeira}, {Terrett}, {Tingley}, {Trager}, {Turon}, {Ulla}, {Utrilla},
  {Valentini}, {van Elteren}, {Van Hemelryck}, {van Leeuwen}, {Varadi},
  {Vecchiato}, {Veljanoski}, {Via}, {Vicente}, {Vogt}, {Voss}, {Votruba},
  {Voutsinas}, {Walmsley}, {Weiler}, {Weingrill}, {Werner}, {Wevers},
  {Whitehead}, {Wyrzykowski}, {Yoldas}, {{\v{Z}}erjal}, {Zucker}, {Zurbach},
  {Zwitter}, {Alecu}, {Allen}, {Allende Prieto}, {Amorim},
  {Anglada-Escud{\'e}}, {Arsenijevic}, {Azaz}, {Balm}, {Beck}, {Bernstein},
  {Bigot}, {Bijaoui}, {Blasco}, {Bonfigli}, {Bono}, {Boudreault}, {Bressan},
  {Brown}, {Brunet}, {Bunclark}, {Buonanno}, {Butkevich}, {Carret}, {Carrion},
  {Chemin}, {Ch{\'e}reau}, {Corcione}, {Darmigny}, {de Boer}, {de Teodoro}, {de
  Zeeuw}, {Delle Luche}, {Domingues}, {Dubath}, {Fodor}, {Fr{\'e}zouls},
  {Fries}, {Fustes}, {Fyfe}, {Gallardo}, {Gallegos}, {Gardiol}, {Gebran},
  {Gomboc}, {G{\'o}mez}, {Grux}, {Gueguen}, {Heyrovsky}, {Hoar}, {Iannicola},
  {Isasi Parache}, {Janotto}, {Joliet}, {Jonckheere}, {Keil}, {Kim},
  {Klagyivik}, {Klar}, {Knude}, {Kochukhov}, {Kolka}, {Kos}, {Kutka}, {Lainey},
  {LeBouquin}, {Liu}, {Loreggia}, {Makarov}, {Marseille}, {Martayan},
  {Martinez-Rubi}, {Massart}, {Meynadier}, {Mignot}, {Munari}, {Nguyen},
  {Nordlander}, {Ocvirk}, {O'Flaherty}, {Olias Sanz}, {Ortiz}, {Osorio},
  {Oszkiewicz}, {Ouzounis}, {Palmer}, {Park}, {Pasquato}, {Peltzer}, {Peralta},
  {P{\'e}turaud}, {Pieniluoma}, {Pigozzi}, {Poels}, {Prat}, {Prod'homme},
  {Raison}, {Rebordao}, {Risquez}, {Rocca-Volmerange}, {Rosen}, {Ruiz-Fuertes},
  {Russo}, {Sembay}, {Serraller Vizcaino}, {Short}, {Siebert}, {Silva},
  {Sinachopoulos}, {Slezak}, {Soffel}, {Sosnowska}, {Strai{\v{z}}ys}, {ter
  Linden}, {Terrell}, {Theil}, {Tiede}, {Troisi}, {Tsalmantza}, {Tur},
  {Vaccari}, {Vachier}, {Valles}, {Van Hamme}, {Veltz}, {Virtanen}, {Wallut},
  {Wichmann}, {Wilkinson}, {Ziaeepour}, \& {Zschocke}}]{GAIA2016overview}
{Gaia Collaboration}, {Prusti}, T., {de Bruijne}, J.~H.~J., {et~al.} 2016,
  \aap, 595, A1

\bibitem[{{Galli} {et~al.}(2021){Galli}, {Bouy}, {Olivares}, {Miret-Roig},
  {Sarro}, {Barrado}, {Berihuete}, {Bertin}, \& {Cuillandre}}]{Galli2021}
{Galli}, P.~A.~B., {Bouy}, H., {Olivares}, J., {et~al.} 2021, \aap, 646, A46

\bibitem[{{Guilloteau} {et~al.}(2016){Guilloteau}, {Pi{\'e}tu}, {Chapillon},
  {Di Folco}, {Dutrey}, {Henning}, {Semenov}, {Birnstiel}, \&
  {Grosso}}]{guilloteau2016}
{Guilloteau}, S., {Pi{\'e}tu}, V., {Chapillon}, E., {et~al.} 2016, \aap, 586,
  L1

\bibitem[{{Honda} {et~al.}(2009){Honda}, {Inoue}, {Fukagawa}, {Oka},
  {Nakamoto}, {Ishii}, {Terada}, {Takato}, {Kawakita}, {Okamoto}, {Shibai},
  {Tamura}, {Kudo}, \& {Itoh}}]{Honda2009}
{Honda}, M., {Inoue}, A.~K., {Fukagawa}, M., {et~al.} 2009, \apjl, 690, L110

\bibitem[{Hunter(2007)}]{matplotlib}
Hunter, J.~D. 2007, Computing in Science \& Engineering, 9, 90

\bibitem[{{Kaeufer} {et~al.}(2023){Kaeufer}, {Woitke}, {Min}, {Kamp}, \&
  {Pinte}}]{kaeufer2023}
{Kaeufer}, T., {Woitke}, P., {Min}, M., {Kamp}, I., \& {Pinte}, C. 2023, arXiv
  e-prints, arXiv:2302.04629

\bibitem[{{Law} {et~al.}(2021){Law}, {Teague}, {Loomis}, {Bae}, {{\"O}berg},
  {Czekala}, {Andrews}, {Aikawa}, {Alarc{\'o}n}, {Bergin}, {Bergner}, {Booth},
  {Bosman}, {Calahan}, {Cataldi}, {Cleeves}, {Furuya}, {Guzm{\'a}n}, {Huang},
  {Ilee}, {Le Gal}, {Liu}, {Long}, {M{\'e}nard}, {Nomura}, {P{\'e}rez}, {Qi},
  {Schwarz}, {Soto}, {Tsukagoshi}, {Yamato}, {van't Hoff}, {Walsh}, {Wilner},
  \& {Zhang}}]{Law21_MAPSIV}
{Law}, C.~J., {Teague}, R., {Loomis}, R.~A., {et~al.} 2021, \apjs, 257, 4

\bibitem[{{Leemker} {et~al.}(2022){Leemker}, {Booth}, {van Dishoeck},
  {P{\'e}rez-S{\'a}nchez}, {Szul{\'a}gyi}, {Bosman}, {Bruderer}, {Facchini},
  {Hogerheijde}, {Paneque-Carre{\~n}o}, \& {Sturm}}]{Leemker2022}
{Leemker}, M., {Booth}, A.~S., {van Dishoeck}, E.~F., {et~al.} 2022, \aap, 663,
  A23

\bibitem[{{Lesur} {et~al.}(2022){Lesur}, {Ercolano}, {Flock}, {Lin}, {Yang},
  {Barranco}, {Benitez-Llambay}, {Goodman}, {Johansen}, {Klahr}, {Laibe},
  {Lyra}, {Marcus}, {Nelson}, {Squire}, {Simon}, {Turner}, {Umurhan}, \&
  {Youdin}}]{lesur2022}
{Lesur}, G., {Ercolano}, B., {Flock}, M., {et~al.} 2022, arXiv e-prints,
  arXiv:2203.09821

\bibitem[{{Lesur} \& {Latter}(2016)}]{Lesur2016}
{Lesur}, G. R.~J. \& {Latter}, H. 2016, \mnras, 462, 4549

\bibitem[{{Luhman}(2007)}]{Luhman2007}
{Luhman}, K.~L. 2007, \apjs, 173, 104

\bibitem[{{Lynden-Bell} \& {Pringle}(1974)}]{lynden-bell1974}
{Lynden-Bell}, D. \& {Pringle}, J.~E. 1974, \mnras, 168, 603

\bibitem[{{Madlener} {et~al.}(2012){Madlener}, {Wolf}, {Dutrey}, \&
  {Guilloteau}}]{Madlener2012}
{Madlener}, D., {Wolf}, S., {Dutrey}, A., \& {Guilloteau}, S. 2012, \aap, 543,
  A81

\bibitem[{{Manara} {et~al.}(2022){Manara}, {Ansdell}, {Rosotti}, {Hughes},
  {Armitage}, {Lodato}, \& {Williams}}]{manara2022_ppvii}
{Manara}, C.~F., {Ansdell}, M., {Rosotti}, G.~P., {et~al.} 2022, arXiv
  e-prints, arXiv:2203.09930

\bibitem[{{Manoj} {et~al.}(2011){Manoj}, {Kim}, {Furlan}, {McClure}, {Luhman},
  {Watson}, {Espaillat}, {Calvet}, {Najita}, {D'Alessio}, {Adame}, {Sargent},
  {Forrest}, {Bohac}, {Green}, \& {Arnold}}]{Manoj2011}
{Manoj}, P., {Kim}, K.~H., {Furlan}, E., {et~al.} 2011, \apjs, 193, 11

\bibitem[{{Marton} {et~al.}(2017){Marton}, {Calzoletti}, {Perez Garcia},
  {Kiss}, {Paladini}, {Altieri}, {Sanchez Portal}, {Kidger}, \& {the Herschel
  Point Source Catalogue Working Group}}]{Marton2017_pacs_psc}
{Marton}, G., {Calzoletti}, L., {Perez Garcia}, A.~M., {et~al.} 2017, arXiv
  e-prints, arXiv:1705.05693

\bibitem[{{McClure}(2009)}]{McClure2009}
{McClure}, M. 2009, \apjl, 693, L81

\bibitem[{{McClure} {et~al.}(2010){McClure}, {Furlan}, {Manoj}, {Luhman},
  {Watson}, {Forrest}, {Espaillat}, {Calvet}, {D'Alessio}, {Sargent}, {Tobin},
  \& {Chiang}}]{McClure2010}
{McClure}, M.~K., {Furlan}, E., {Manoj}, P., {et~al.} 2010, \apjs, 188, 75

\bibitem[{{McMullin} {et~al.}(2007){McMullin}, {Waters}, {Schiebel}, {Young},
  \& {Golap}}]{McMullin07}
{McMullin}, J.~P., {Waters}, B., {Schiebel}, D., {Young}, W., \& {Golap}, K.
  2007, in Astronomical Society of the Pacific Conference Series, Vol. 376,
  Astronomical Data Analysis Software and Systems XVI, ed. R.~A. {Shaw},
  F.~{Hill}, \& D.~J. {Bell}, 127

\bibitem[{{Min} {et~al.}(2005){Min}, {Hovenier}, \& {de Koter}}]{Min2005DHS}
{Min}, M., {Hovenier}, J.~W., \& {de Koter}, A. 2005, \aap, 432, 909

\bibitem[{{Olofsson} {et~al.}(2009){Olofsson}, {Augereau}, {van Dishoeck},
  {Mer{\'\i}n}, {Lahuis}, {Kessler-Silacci}, {Dullemond}, {Oliveira}, {Blake},
  {Boogert}, {Brown}, {Evans}, {Geers}, {Knez}, {Monin}, \&
  {Pontoppidan}}]{Olofsson2009}
{Olofsson}, J., {Augereau}, J.~C., {van Dishoeck}, E.~F., {et~al.} 2009, \aap,
  507, 327

\bibitem[{{Padgett} {et~al.}(1999){Padgett}, {Brandner}, {Stapelfeldt},
  {Strom}, {Terebey}, \& {Koerner}}]{Padgett1999}
{Padgett}, D.~L., {Brandner}, W., {Stapelfeldt}, K.~R., {et~al.} 1999, \aj,
  117, 1490

\bibitem[{{Pickles}(1998)}]{Pickles1998}
{Pickles}, A.~J. 1998, \pasp, 110, 863

\bibitem[{{Pinte} {et~al.}(2016){Pinte}, {Dent}, {M{\'e}nard}, {Hales}, {Hill},
  {Cortes}, \& {de Gregorio-Monsalvo}}]{pinte2016}
{Pinte}, C., {Dent}, W.~R.~F., {M{\'e}nard}, F., {et~al.} 2016, \apj, 816, 25

\bibitem[{{Podio} {et~al.}(2020){Podio}, {Garufi}, {Codella}, {Fedele},
  {Bianchi}, {Bacciotti}, {Ceccarelli}, {Favre}, {Mercimek}, {Rygl}, \&
  {Testi}}]{Podio2020}
{Podio}, L., {Garufi}, A., {Codella}, C., {et~al.} 2020, \aap, 642, L7

\bibitem[{{Pontoppidan} {et~al.}(2005){Pontoppidan}, {Dullemond}, {van
  Dishoeck}, {Blake}, {Boogert}, {Evans}, {Kessler-Silacci}, \&
  {Lahuis}}]{Pontoppidan2005}
{Pontoppidan}, K.~M., {Dullemond}, C.~P., {van Dishoeck}, E.~F., {et~al.} 2005,
  \apj, 622, 463

\bibitem[{{Pontoppidan} {et~al.}(2007){Pontoppidan}, {Stapelfeldt}, {Blake},
  {van Dishoeck}, \& {Dullemond}}]{Pontoppidan2007}
{Pontoppidan}, K.~M., {Stapelfeldt}, K.~R., {Blake}, G.~A., {van Dishoeck},
  E.~F., \& {Dullemond}, C.~P. 2007, \apjl, 658, L111

\bibitem[{{Robberto} {et~al.}(2012){Robberto}, {Spina}, {Da Rio}, {Apai},
  {Pascucci}, {Ricci}, {Goddi}, {Testi}, {Palla}, \&
  {Bacciotti}}]{Robberto2012}
{Robberto}, M., {Spina}, L., {Da Rio}, N., {et~al.} 2012, \aj, 144, 83

\bibitem[{{Sauter} {et~al.}(2009){Sauter}, {Wolf}, {Launhardt}, {Padgett},
  {Stapelfeldt}, {Pinte}, {Duch{\^e}ne}, {M{\'e}nard}, {McCabe}, {Pontoppidan},
  {Dunham}, {Bourke}, \& {Chen}}]{Sauter2009}
{Sauter}, J., {Wolf}, S., {Launhardt}, R., {et~al.} 2009, \aap, 505, 1167

\bibitem[{{Siess} {et~al.}(2000){Siess}, {Dufour}, \& {Forestini}}]{siess2000}
{Siess}, L., {Dufour}, E., \& {Forestini}, M. 2000, \aap, 358, 593

\bibitem[{{Stapelfeldt} {et~al.}(2014){Stapelfeldt}, {Duch{\^e}ne}, {Perrin},
  {Wolff}, {Krist}, {Padgett}, {M{\'e}nard}, \& {Pinte}}]{Stapelfeldt2014}
{Stapelfeldt}, K.~R., {Duch{\^e}ne}, G., {Perrin}, M., {et~al.} 2014, in
  Exploring the Formation and Evolution of Planetary Systems, ed. M.~{Booth},
  B.~C. {Matthews}, \& J.~R. {Graham}, Vol. 299, 99--103

\bibitem[{{Sturm} {et~al.}(2023{\natexlab{a}}){Sturm}, {Booth}, {McClure},
  {Leemker}, \& {van Dishoeck}}]{Sturm2022lkca15}
{Sturm}, J.~A., {Booth}, A.~S., {McClure}, M.~K., {Leemker}, M., \& {van
  Dishoeck}, E.~F. 2023{\natexlab{a}}, \aap, 670, A12

\bibitem[{{Sturm} {et~al.}(2023{\natexlab{b}}){Sturm}, {McClure}, {Bergner},
  {Harsono}, {Dartois}, {Drozdovskaya}, {Ioppolo}, {{\"O}berg}, {Law},
  {Palumbo}, {Pendleton}, {Rocha}, {Terada}, \& {Urso}}]{PaperII}
{Sturm}, J.~A., {McClure}, M.~K., {Bergner}, J.~B., {et~al.}
  2023{\natexlab{b}}, arXiv e-prints, arXiv:2305.02355

\bibitem[{{Tabone} {et~al.}(2020){Tabone}, {Cabrit}, {Pineau des For{\^e}ts},
  {Ferreira}, {Gusdorf}, {Podio}, {Bianchi}, {Chapillon}, {Codella}, \&
  {Gueth}}]{Tabone2020}
{Tabone}, B., {Cabrit}, S., {Pineau des For{\^e}ts}, G., {et~al.} 2020, \aap,
  640, A82

\bibitem[{{Tabone} {et~al.}(2022){Tabone}, {Rosotti}, {Lodato}, {Armitage},
  {Cridland}, \& {van Dishoeck}}]{Tabone2022}
{Tabone}, B., {Rosotti}, G.~P., {Lodato}, G., {et~al.} 2022, \mnras, 512, L74

\bibitem[{{Tazaki} {et~al.}(2021){Tazaki}, {Murakawa}, {Muto}, {Honda}, \&
  {Inoue}}]{Tazaki2021}
{Tazaki}, R., {Murakawa}, K., {Muto}, T., {Honda}, M., \& {Inoue}, A.~K. 2021,
  \apj, 921, 173

\bibitem[{{Teague}(2019)}]{Teague19_eddy}
{Teague}, R. 2019, The Journal of Open Source Software, 4, 1220

\bibitem[{{Teague} \& {Foreman-Mackey}(2018)}]{Teague18}
{Teague}, R. \& {Foreman-Mackey}, D. 2018, Research Notes of the American
  Astronomical Society, 2, 173

\bibitem[{{Teague} {et~al.}(2020){Teague}, {Jankovic}, {Haworth}, {Qi}, \&
  {Ilee}}]{Teague2020}
{Teague}, R., {Jankovic}, M.~R., {Haworth}, T.~J., {Qi}, C., \& {Ilee}, J.~D.
  2020, \mnras, 495, 451

\bibitem[{{Trapman} {et~al.}(2019){Trapman}, {Facchini}, {Hogerheijde}, {van
  Dishoeck}, \& {Bruderer}}]{Trapman2019}
{Trapman}, L., {Facchini}, S., {Hogerheijde}, M.~R., {van Dishoeck}, E.~F., \&
  {Bruderer}, S. 2019, \aap, 629, A79

\bibitem[{{van der Marel}(2022)}]{vandermarel2022}
{van der Marel}, N. 2022, arXiv e-prints, arXiv:2210.05539

\bibitem[{{van der Marel} \& {Mulders}(2021)}]{vandermarel2021_deomgraphics}
{van der Marel}, N. \& {Mulders}, G.~D. 2021, \aj, 162, 28

\bibitem[{van~der Walt {et~al.}(2011)van~der Walt, Colbert, \&
  Varoquaux}]{numpy}
van~der Walt, S., Colbert, S.~C., \& Varoquaux, G. 2011, Computing in Science
  \& Engineering, 13, 22

\bibitem[{{Villenave} {et~al.}(2019){Villenave}, {Benisty}, {Dent},
  {M{\'e}nard}, {Garufi}, {Ginski}, {Pinilla}, {Pinte}, {Williams}, {de Boer},
  {Morino}, {Fukagawa}, {Dominik}, {Flock}, {Henning}, {Juh{\'a}sz}, {Keppler},
  {Muro-Arena}, {Olofsson}, {P{\'e}rez}, {van der Plas}, {Zurlo}, {Carle},
  {Feautrier}, {Pavlov}, {Pragt}, {Ramos}, {Sauvage}, {Stadler}, \&
  {Weber}}]{Villenave2019}
{Villenave}, M., {Benisty}, M., {Dent}, W.~R.~F., {et~al.} 2019, \aap, 624, A7

\bibitem[{{Villenave} {et~al.}(2020){Villenave}, {M{\'e}nard}, {Dent},
  {Duch{\^e}ne}, {Stapelfeldt}, {Benisty}, {Boehler}, {van der Plas}, {Pinte},
  {Telkamp}, {Wolff}, {Flores}, {Lesur}, {Louvet}, {Riols}, {Dougados},
  {Williams}, \& {Padgett}}]{Villenave2020}
{Villenave}, M., {M{\'e}nard}, F., {Dent}, W.~R.~F., {et~al.} 2020, \aap, 642,
  A164

\bibitem[{{Villenave} {et~al.}(2022){Villenave}, {Stapelfeldt}, {Duch{\^e}ne},
  {M{\'e}nard}, {Lambrechts}, {Sierra}, {Flores}, {Dent}, {Wolff}, {Ribas},
  {Benisty}, {Cuello}, \& {Pinte}}]{villenave2022}
{Villenave}, M., {Stapelfeldt}, K.~R., {Duch{\^e}ne}, G., {et~al.} 2022, \apj,
  930, 11

\bibitem[{{Walsh} {et~al.}(2015){Walsh}, {Nomura}, \& {van
  Dishoeck}}]{Walsh2015}
{Walsh}, C., {Nomura}, H., \& {van Dishoeck}, E. 2015, \aap, 582, A88

\bibitem[{{Weingartner} \& {Draine}(2001)}]{weingartner2001}
{Weingartner}, J.~C. \& {Draine}, B.~T. 2001, \apj, 548, 296

\bibitem[{{Windmark} {et~al.}(2012){Windmark}, {Birnstiel}, {G{\"u}ttler},
  {Blum}, {Dullemond}, \& {Henning}}]{Windmark2012}
{Windmark}, F., {Birnstiel}, T., {G{\"u}ttler}, C., {et~al.} 2012, \aap, 540,
  A73

\bibitem[{{Winston} {et~al.}(2012){Winston}, {Cox}, {Prusti}, {Mer{\'\i}n},
  {Ribas}, {Royer}, {Vavrek}, {Puga}, {Andr{\'e}}, {Men'shchikov},
  {K{\"o}nyves}, {K{\'o}sp{\'a}l}, {Alves de Oliveira}, {Pilbratt}, \&
  {Waelkens}}]{Winston2012}
{Winston}, E., {Cox}, N.~L.~J., {Prusti}, T., {et~al.} 2012, \aap, 545, A145

\bibitem[{{Wolff} {et~al.}(2021){Wolff}, {Duch{\^e}ne}, {Stapelfeldt},
  {M{\'e}nard}, {Flores}, {Padgett}, {Pinte}, {Villenave}, {van der Plas}, \&
  {Perrin}}]{wolff2021}
{Wolff}, S.~G., {Duch{\^e}ne}, G., {Stapelfeldt}, K.~R., {et~al.} 2021, \aj,
  161, 238

\bibitem[{{Wolff} {et~al.}(2017){Wolff}, {Perrin}, {Stapelfeldt},
  {Duch{\^e}ne}, {M{\'e}nard}, {Padgett}, {Pinte}, {Pueyo}, \&
  {Fischer}}]{wolff2017}
{Wolff}, S.~G., {Perrin}, M.~D., {Stapelfeldt}, K., {et~al.} 2017, \apj, 851,
  56

\bibitem[{{Zsidi} {et~al.}(2022){Zsidi}, {Manara}, {K{\'o}sp{\'a}l}, {Hussain},
  {{\'A}brah{\'a}m}, {Alecian}, {B{\'o}di}, {P{\'a}l}, \& {Sarkis}}]{zsidi2022}
{Zsidi}, G., {Manara}, C.~F., {K{\'o}sp{\'a}l}, {\'A}., {et~al.} 2022, \aap,
  660, A108

\end{thebibliography}

\begin{appendix}
\section{posterior distributions continuum fits}
\begin{figure*}
    \includegraphics[width = \textwidth]{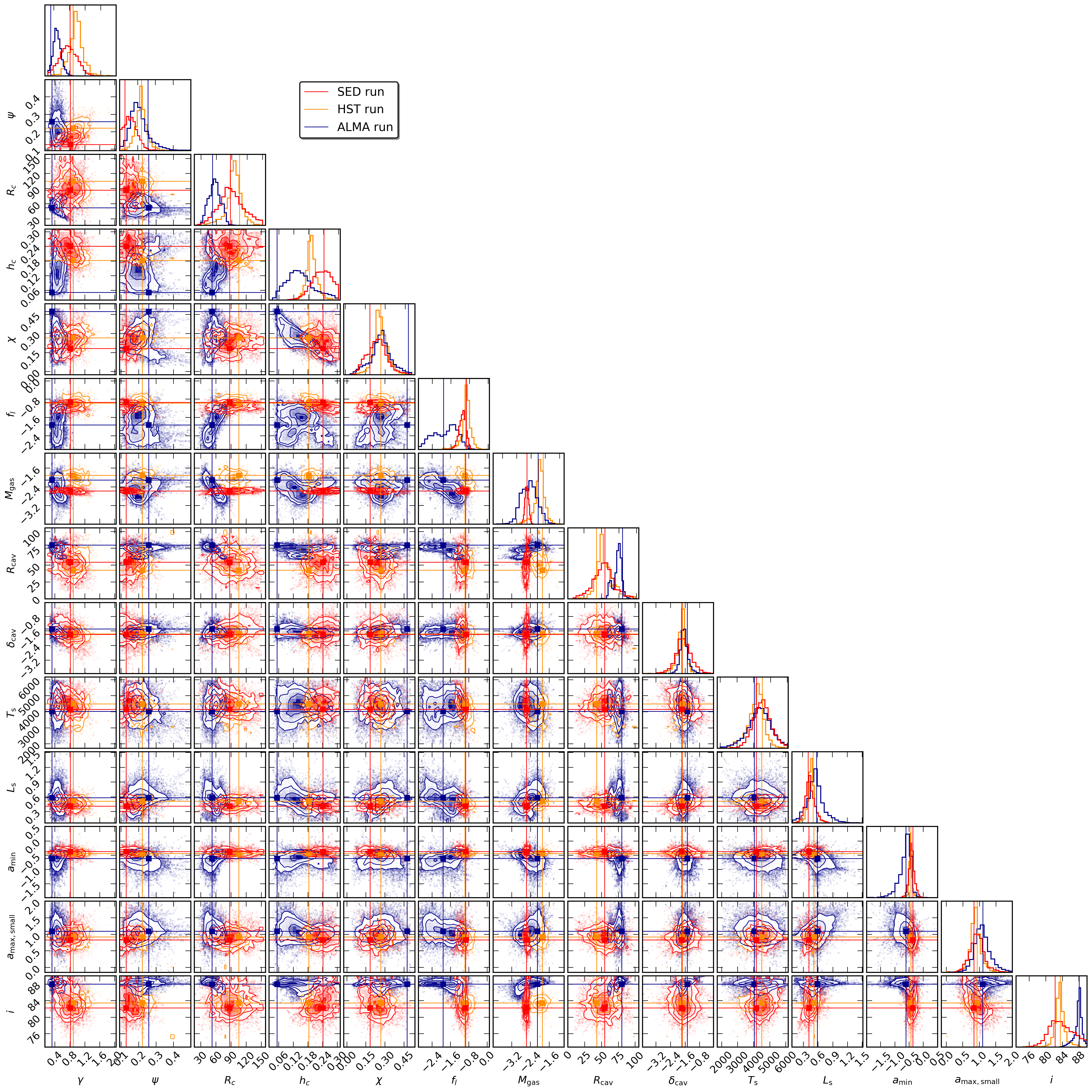}
    \caption{Posterior distributions of the three different MCMC runs. The SED run is shown in red, the HST run in orange and the ALMA run in blue. The median value for each parameter and the 16\% and 84\% confidence intervals are given in Table~\ref{tab:prior_results}.}
    \label{fig:posteriors}
\end{figure*}

\begin{figure*}
    \includegraphics[width = \textwidth]{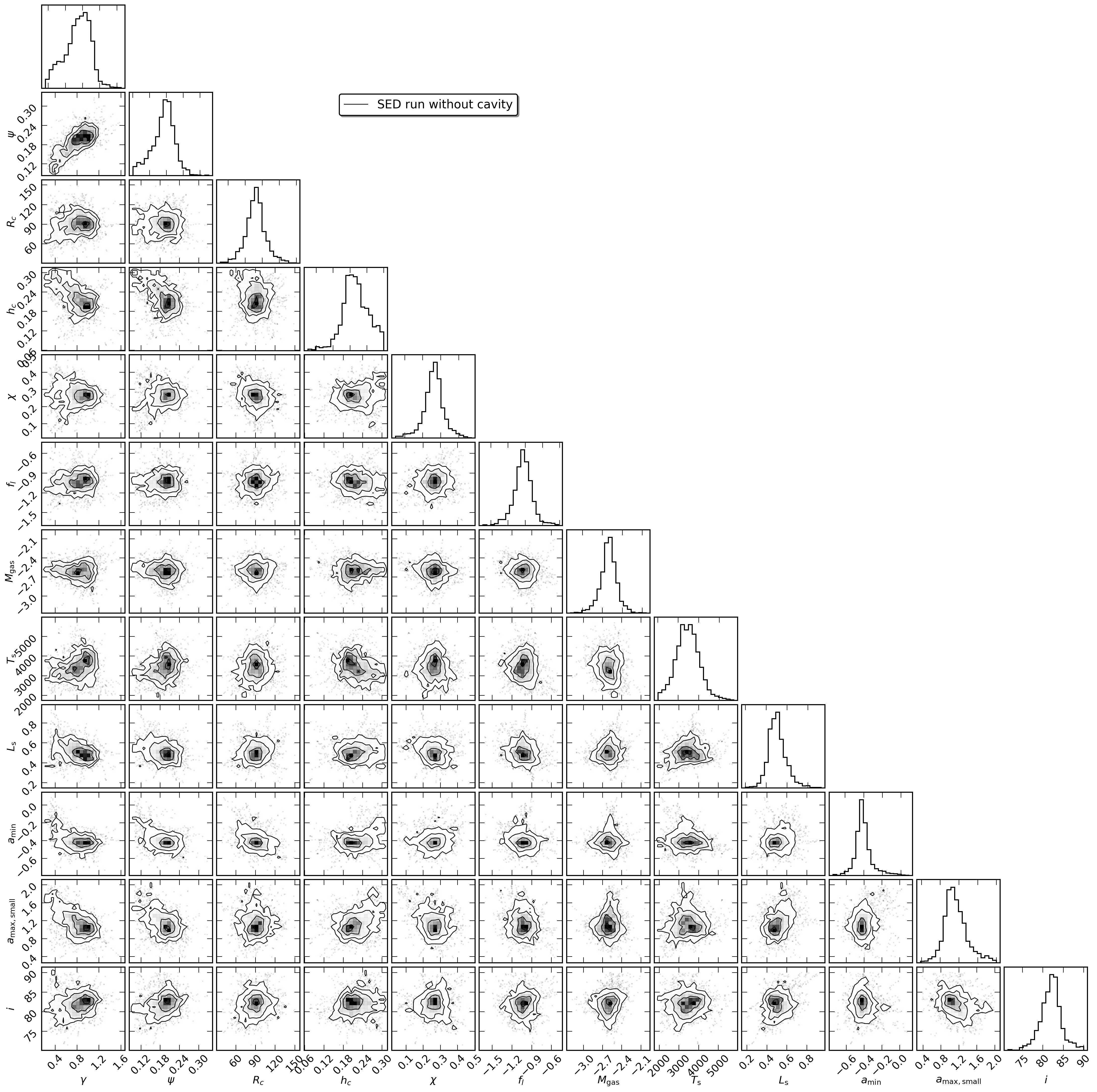}
    \caption{Posterior distribution of the SED run without a cavity. The median value for each parameter and the 16\% and 84\% confidence intervals are given in Table~\ref{tab:prior_results}.}
    \label{fig:posterior_nocav}
\end{figure*}

In Figure \ref{fig:posteriors}, we present the posterior distribution of the three independent runs using the SED, the resolved scattered light observation and the resolved millimeter continuum observations. In Figure \ref{fig:posterior_nocav}, we present the posterior distribution of the SED run without a cavity, the physical parameters of which are overall in good agreement with the SED run with a cavity.

\label{app:posteriors}

\end{appendix}
\end{document}